\documentclass{article}

\usepackage{arxiv}

\usepackage[utf8]{inputenc} 
\usepackage[T1]{fontenc}    
\usepackage{hyperref}       
\usepackage{url}            
\usepackage{booktabs}       
\usepackage{amsfonts}       
\usepackage{nicefrac}       
\usepackage{microtype}      
\usepackage{lipsum}
\usepackage{graphicx}
\graphicspath{ {./images/} }

\usepackage{caption}
\usepackage{subcaption}
\usepackage{algorithm}
\usepackage{algcompatible}
\usepackage{multirow}
\usepackage{textcomp, gensymb}
\usepackage{url}
\usepackage{makecell}

\usepackage[utf8]{inputenc}

\usepackage{siunitx}
\usepackage{booktabs}
\usepackage{graphicx}
\usepackage{amsmath}
\usepackage{siunitx}
\usepackage{stfloats}
\usepackage{xcolor}
\usepackage{graphicx}
\usepackage{colortbl}
\usepackage{multirow}

\usepackage{makecell}

\usepackage{xstring}
\let\oldtexttt\texttt
\renewcommand{\texttt}[1]{\oldtexttt{\StrSubstitute[0]{#1}{.}{.\allowbreak}}}

\usepackage{tikz}
\usepackage{pgfplots}
\usetikzlibrary{fillbetween}
\usepgfplotslibrary{dateplot}
\pgfplotsset{compat=1.17}

\usepackage[eulergreek]{sansmath}
\pgfplotsset{
every x tick label/.append style={font=\scriptsize\sansmath\sffamily},
every y tick label/.append style={font=\scriptsize\sansmath\sffamily},
every axis label/.append style={font=\small\slshape},
every axis legend/.append style={font=\small},
}

\definecolor{bargreen}{rgb}{0.72, 0.88, 0.65}

\definecolor{sparkblue}{rgb}{0.46, 0.76, 0.71}

\makeatletter
\newcommand\resetstackedplots{
\makeatletter
\pgfplots@stacked@isfirstplottrue
\makeatother
\addplot [forget plot,draw=none] coordinates{(0,0) (1,0) (2,0) (3,0) (4,0) (5,0) (6,0) (7,0)};
}

\usepackage{listings}
\usepackage{courier}
\definecolor{codegreen}{rgb}{0,0.6,0}

\title{Does Big Data Require Complex Systems?\\A Performance Comparison Between Spark and Unicage~Shell Scripts}

\date{} 					

\author{
  Duarte M.Nascimento, Miguel L. Pardal \\
  INESC-ID, Instituto Superior T\'{e}cnico, Universidade de Lisboa\\
  Lisbon, Portugal\\
  \texttt{\{duarte.miguel,miguel.pardal\}@tecnico.ulisboa.pt} \\
   \And
  Miguel Ferreira\\
  Unicage Europe\\
  Lisbon, Portugal\\
  \texttt{miguel.ferreira@unicage.com} \\
}

\begin{document}

\maketitle

\begin{abstract}
The paradigm of big data is characterized by the need to collect and process data sets of great volume, arriving at the systems with great velocity, in a variety of formats. 
Spark is a widely used big data processing system that can be integrated with Hadoop to provide powerful abstractions to developers, such as distributed storage through HDFS and resource management through YARN.
When all the required configurations are made, Spark can also provide quality attributes, such as scalability, fault tolerance, and security.
However, all of these benefits come at the cost of complexity, with high memory requirements, and additional latency in processing.
An alternative approach is to use a lean software stack, like Unicage, that delegates most control back to the developer.

In this work we evaluated the performance of  big data processing with Spark versus Unicage, in a cluster environment hosted in the IBM Cloud.
Two sets of experiments were performed: batch processing of unstructured data sets, and query processing of structured data sets.
The input data sets were of significant size, ranging from 64\,GB to 8192\,GB in volume.
The results show that the performance of  Unicage scripts is superior to Spark for search workloads like \emph{grep} and \emph{select}, but that the abstractions of distributed storage and resource management from the Hadoop stack enable Spark to execute workloads with inter-record dependencies, such as \emph{sort} and \emph{join}, with correct outputs.
\end{abstract}

\section{Introduction}
\label{sec:introduction}

The paradigm of big data has been growing with the rise of the Internet of Things (IoT). Sensors and actuators now join the already established sources of data, to produce even larger quantities, at faster speeds. Juniper Research estimates an increase of 137\% in the number of IoT connections from 2020 to 2024~\cite{rothmuller2020iot}. Statista estimates an increase of 129\% in the amount of generated data from 2020 to 2024~\cite{holst2021volume}.

Big data is characterized by a set of \emph{V properties}: \emph{volume}, \emph{velocity} and \emph{variety} of sources and formats characterize the dimensions of data, while \emph{veracity} and \emph{value} characterize its quality. Common data becomes big data once handling these attributes transcends the capabilities of a single machine or of a small cluster~\cite{demauro2016formal}. The cloud is a resourceful environment for the deployment of big data systems and the  processing of workloads in production. Big data systems seek to handle the dimensions of data sets while preserving their quality. \emph{Hadoop MapReduce} was the popular big data system of yore, but as the paradigm evolved, so did the necessity for more capable systems, like \emph{Spark}\footnote{https://spark.apache.org/}.
Most recent big data systems offer a plethora of abstractions to allow the developer to focus strictly on the data and how to deal with it. These abstractions include distributed storage (e.g., HDFS in the Hadoop software stack) and distributed resource management (e.g., YARN in the Hadoop software stack). Additionally, 
these systems seek to optimize a variety of quality attributes including performance, scalability, fault tolerance, security, usability and manageability. As a result, the systems harboring these responsibilities have inevitably complex software stacks and are hereafter addressed as \emph{complex big data systems}. The complexity of these software stacks affects the performance of the big data processing workloads, with additional latency. Furthermore, this complexity hinders the deployment in low-powered devices with low hardware specifications, such as IoT devices.

As an alternative to the shortcomings of complex big data systems, \emph{lean big data systems} are characterized by an absence of many abstraction layers, and put the responsibility to implement the needed mechanisms mostly on the developer. This way, these solutions can be tailored to the needs of the use-case and optimized only for the required set of quality attributes. Unicage\footnote{https://unicage.eu/} enables the implementation of lean big data systems through shell scripting. 

\begin{figure}[ht]
\includegraphics[width=0.75\columnwidth]{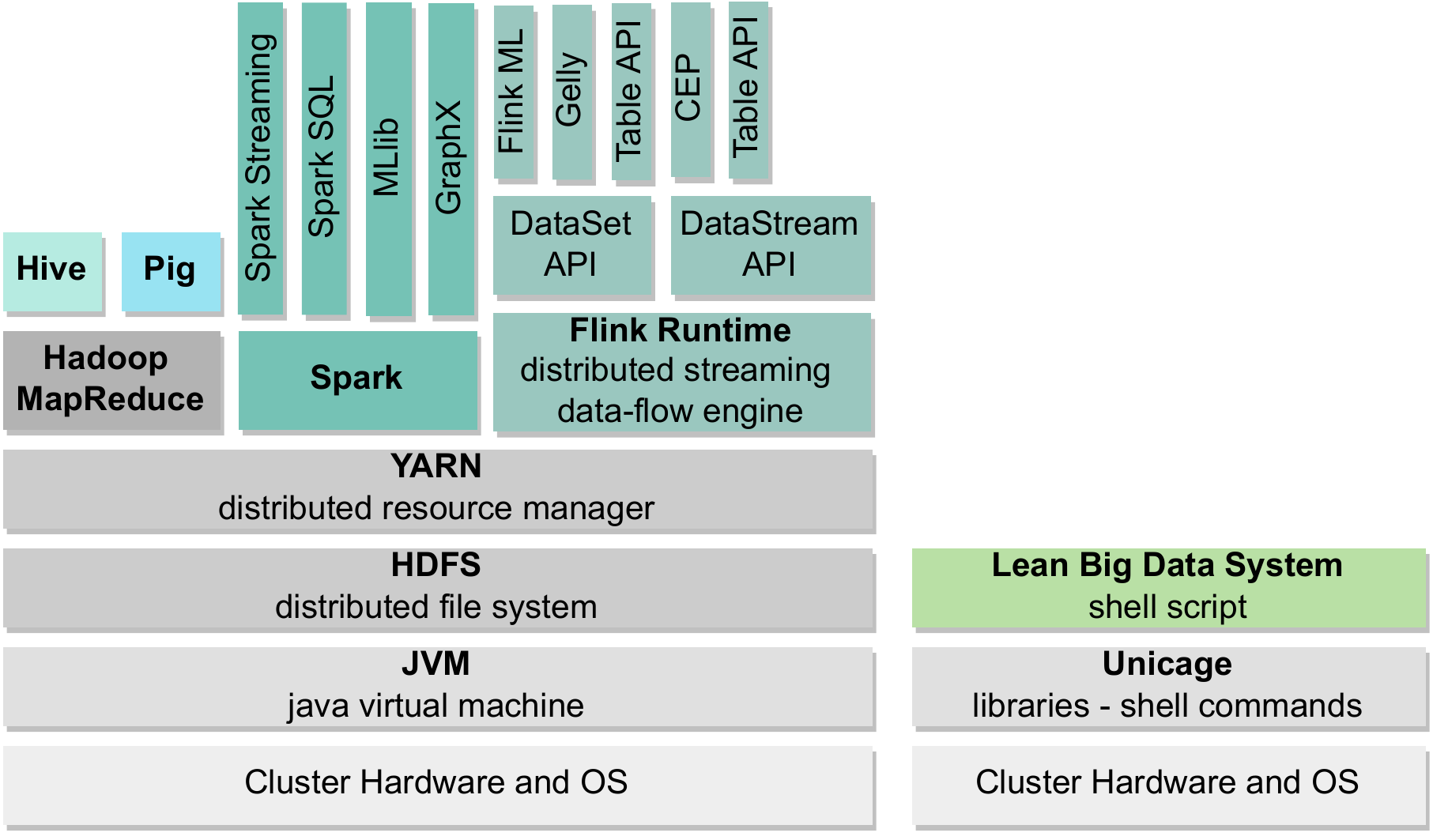}
\centering
\caption[Software stacks of complex big data systems provided by the Apache Software Foundation and of a lean big data system provided by Unicage.]{Software stacks of complex big data systems provided by the Apache Software Foundation (left); and of a lean big data system implemented with Unicage (right).}
\label{fig:software-stacks}
\end{figure}

Figure~\ref{fig:software-stacks} illustrates the software stack difference between complex and lean big data systems. 

This study presents the first benchmark that compares complex and lean big data systems directly in a cloud environment, and is focused strictly on the performance evaluation of data loading, batch processing and query processing workloads. Spark was benchmarked as the main representative of complex big data systems, as it is widely used and has been, in recent years, one of the most popular projects hosted by the Apache Software Foundation~\cite{sally2022apachebythedigits}. Hadoop MapReduce and Hive (over MapReduce) are not as widely used nowadays, but were also benchmarked for historical reference, and as tiebreakers for the verification of results. Unicage was benchmarked as the representative of lean big data systems.

\section{Background}
\label{sec:background}

This section provides a brief description of the big data systems benchmarked in this study.

\subsection{Complex Big Data Systems}

The Hadoop stack runs on top of the Java Virtual Machine (JVM), featuring HDFS~\cite{shvachko2010thehd} and YARN~\cite{vavilapalli2013apache} as the underlying distributed file system and distributed resource manager, respectively. Hadoop MapReduce~\cite{dean2008mapreduce} is a big data system for batch processing. Hive~\cite{thusoo2009hive,thusoo2010hive} originally enabled query processing on top of MapReduce, but has been extended to use Spark as the underlying processing engine instead. Spark~\cite{zaharia2010spark} and Flink~\cite{carbone2015apache} are multipurpose big data systems that can use the distributed storage and resource management capabilities of the Hadoop stack.

\subsubsection{MapReduce}

MapReduce is a programming model designed for batch processing of large volumes of data, represented as key/va\-lue pairs. The typical MapReduce application is comprised of two user-defined operations: the \emph{Mapper} operation that processes the input key/value pairs and produces intermediate key/value pairs; and the \emph{Reducer} operation that processes the intermediate key/value pairs output by the Mapper and produces output key/value pairs. A \emph{Split} operation often precedes the Map operation to split the input data set in records of a user-defined size, to be processed by individual Mappers. A \emph{Shuffle} operation often succeeds the Map operation to group the intermediate key/value pairs output by the Mappers in logic key/value pairs, sorted by key, to be processed by individual Reducers. The Split and Shuffle operations allow Mappers and Reducers to be parallelized.

\subsubsection{HDFS}

The Hadoop Distributed File System enables distributed storage across a cluster with multiple nodes, presenting it as a single global namespace to both the user and applications. HDFS provides other quality attributes to the Hadoop software stack, such as fault tolerance through self-managed file block replication, and scalability through a leader/follower architecture. The architecture of HDFS provides these capabilities through one or more leader nodes called \emph{NameNodes}, where file metadata is persisted; and follower nodes called \emph{DataNodes}, where the raw data of files is persisted and replicated. 

One of the shortcomings of HDFS is the latency associated with file block management and lookup. To minimize this, file block size is set to 128\,MB, and the maximum number of blocks per file is set to 10000, both by default. When there is a large number of files whose size is significantly smaller than the set file block size, causing a large number of very small file blocks to exist within the file system, the performance of HDFS is known to deteriorate~\cite{dong2012optimized}.

\subsubsection{YARN}

``Yet Another Resource Negociator'' is the distributed resource manager of the Hadoop software stack, providing load balancing and cluster scalability through the dynamic negotiation of resources to each Hadoop application running in the hosting cluster. YARN enables \emph{multi-tenancy}, allowing multiple logically separated Hadoop applications to share physical resources in a shared environment, each translating to an isolated YARN application. Managed resources include allocated memory, CPU, and disk space. Their negotiation is handled by: a \emph{ResourceManager} service that runs in the NameNode and is responsible for the resource allocation in the form of bundled application containers; one \emph{NodeManager} service running in each DataNode, responsible for updating the ResourceManager with the status of the corresponding DataNode; and one \emph{ApplicationMaster} service running per-application, responsible for the negotiation of resources with the ResourceManager and the execution, management and monitoring of the YARN application containers.

\subsubsection{Hive}

Hive is a data warehousing solution that enables query processing of structured data in Hadoop. Programming in MapReduce is notoriously difficult, so Hive offers an SQL-based querying language called \emph{HiveQL}, whose statements are converted to MapReduce jobs. A structured data set to be processed with HiveQL queries needs to be stored in HDFS, and imported to the Hive database in tabular format, along with its schema. This does not imply copying the data set over to the database, as only the schema and information about the tables is persisted in a metadata repository called \emph{Metastore}. Modern implementations of Hive use Spark as the underlying processing engine, instead of MapReduce.

\subsubsection{Spark}

Oftentimes, Hadoop applications perform iterative algorithms that require MapReduce jobs to load data from disk repeatedly, which is an inefficient use of resources. 
Spark abandons the MapReduce paradigm in favor of partioning data sets into specialized immutable data structures called \emph{resilient distributed data sets} (RDDs)~\cite{zaharia2012resilient}.
Jobs are compiled into direct acyclic graphs of operations to be performed on RDDs. RDDs are partitioned across the nodes of a Spark cluster and their data lineage information is used to rebuild them if nodes fail or partitions are lost. RDD partitions are stored in memory and loaded on demand, providing a significant performance increase when compared with the MapReduce approach. 
Caching may result in memory becoming full, to which Spark reacts by ceasing to cache RDD partitions and instead recomputing them on demand, allowing the job to still proceed but with reduced performance. Alternatively, the caching level can also be configured so RDD partitions that do not fit in memory are stored and read from disk instead.

Spark has a leader/follower architecture that allows its integration with HDFS and YARN. A Spark application runs in the context of a \emph{SparkContext}. A typical Spark cluster features two types of processes: one \emph{Driver} process, responsible for the creation, execution and management of the SparkContext, the scheduling of the direct acyclic graphs and planning of tasks, and the communication with YARN for resource negotiation; and one or more \emph{Executor} processes, responsible for the execution of the tasks planned by the Driver. 

\emph{Cluster mode} enables the Driver process to execute in one of the worker nodes. If Spark is deployed with YARN, the NameNode runs the ResourceManager service, the Driver process runs in one of the DataNodes along with the ApplicationMaster service, and the Executors also run in the DataNodes as YARN application containers. 

The architecture of Spark, combined with the immutable nature of RDDs allowed it to expand beyond its original purpose of batch processing to feature various domain-specific libraries, including \emph{Spark SQL}~\cite{armbrust2015spark} for query processing; \emph{Spark Streaming}~\cite{zaharia2013discretized} and \emph{Spark Structured Streaming}~\cite{armbrust2018structured} for stream processing; \emph{GraphX}~\cite{gonzalez2014graphx} for graph processing; and \emph{MLlib}~\cite{meng2016mllib} for machine learning.

\subsection{Lean Big Data Systems}

Lean big data systems delegate most responsibilities to the developer. Unicage embraces the UNIX fundamentals and offers a set of proprietary commands written in the C language, that can be used alongside typical shell commands to implement data processing pipelines within shell scripts. 
The C language typically uses less memory and consumes less energy than other languages~\cite{pereira2017energy}. 
Unicage features a base library for single-machine data processing called \emph{Tukubai}, and an extension of this library with additional commands for parallel and distributed data processing called \emph{BOA}.

\subsubsection{Unicage Tukubai}

Tukubai features 200 commands designed for single-machine data processing. Table~\ref{tab:unicage-commands} describes some of the most commonly used commands in the library.

\begin{table}[H]
\centering
\begin{tabular}{lc}
\hline
      Command & Description                                                                 \\ \hline
\rowcolor[HTML]{F5F5F5} 
\texttt{msort}  & Performs in-memory sort of records using the merge-sort algorithm.          \\
\texttt{lcnt}   & Counts the number of records in a file.                                     \\
\rowcolor[HTML]{F5F5F5} 
\texttt{dmerge} & \cellcolor[HTML]{F5F5F5}Merges two sorted files based on specified columns. \\
\texttt{self}   & Selects and reorders the columns in a file.                                         \\
\rowcolor[HTML]{F5F5F5} 
\texttt{sm2} & \cellcolor[HTML]{F5F5F5}Aggregates records grouped by specified columns. \\ \hline
\end{tabular}
\caption{Examples of Unicage Tukubai commands.}
\label{tab:unicage-commands}
\end{table}

Ferreira et al.~\cite{ferreira2021smart} demonstrated a use-case with Tukubai, applied to processing data generated from a smart grid. 

\subsubsection{Unicage BOA}

``BigData Oriented Architecture'' features parallel and distributed counterparts for the commonly used commands in the Tukubai library. The typical BOA cluster has a leader/follower architecture, and the followers that partake in the computation are specified per-command. BOA features three types of commands: the \texttt{para-*} commands, aimed at parallel data processing in a single machine; the \texttt{clust-*} commands, aimed at parallel and distributed data processing in a cluster of machines, and are typically issued from the leader node; and the \texttt{distr-*} commands, similar to the \texttt{clust-*} commands, but the leader node does not take part in the computation. Table~\ref{tab:boa-commands} lists some commonly used commands in the BOA library.

\begin{table}[H]
\centering
\begin{tabular}{lccc}
\cline{2-4}
                      & \multicolumn{3}{c}{BOA commands}          \\ \hline
Tukubai base command & \texttt{para-*}      & \texttt{clust-*}      & \texttt{distr-*}      \\ \hline
\rowcolor[HTML]{F5F5F5} 
\texttt{msort}        & \color[HTML]{9B9B9B} n/a  & \texttt{clust-msort}  & \texttt{distr-msort}  \\
\texttt{dmerge}       & \color[HTML]{9B9B9B} n/a & \color[HTML]{9B9B9B} n/a & \texttt{distr-dmerge} \\
\rowcolor[HTML]{F5F5F5} 
\texttt{self}         & \texttt{para-self}   & \texttt{clust-self}   & \texttt{distr-self}   \\
\texttt{sm2}       & \texttt{para-sm2} & \texttt{clust-sm2} & \texttt{distr-sm2} \\ \hline
\end{tabular}
\caption{Examples of Unicage BOA commands.}
\label{tab:boa-commands}
\end{table}

Listing~\ref{listing:unicage} shows a simplified implementation of \emph{wordcount} with a shell script using Unicage commands. 
This script is to be executed from a leader node.
Here, the sub-script to be run by each individual worker is written as a \emph{heredoc} in the \texttt{$\sim$/worker\_script.sh} file. 
This sub-script results in the \texttt{$\sim$/wk\_output} file, with the count of the words in the fraction of the data set of each worker.
It is then distributed and executed in each worker with the \texttt{distr-shell} command. 
Finally, the results of each worker are merged with the \texttt{distr-dmerge} command, which produces a sorted file that is then pipelined into the \texttt{sm2} command to be aggregated, resulting in the \texttt{$\sim$/output} file in the leader node, with two columns - the sorted list of words and the counts of each word.

\noindent
\begin{minipage}{\linewidth}
\hfill
\lstset{emph={ls, grep, sort, head, tr, msort, count, dmerge, sm2, distr, shell, dmerge},emphstyle=\textbf}
\begin{lstlisting}[
language=bash, caption={Shell script of a distributed \emph{wordcount} with Unicage commands.}, label={listing:unicage}, captionpos=b, numbers=left, xleftmargin=1.6em, framexleftmargin=1.5em, commentstyle=\color{codegreen}, basicstyle=\small]
cat >> ~/worker_script.sh <<'EOF'
  tr " " "\n" < ~/dataset.txt |
  msort key=1 | # Tukubai
  count 1 1 | # Tukubai
  sm2 1 1 2 2 > ~/wk_output # Tukubai
EOF

distr-shell $workers ~/worker_script.sh # BOA
distr-dmerge $workers key=1 ~/wk_output | # BOA
sm2 1 1 2 2 > ~/output # Tukubai

\end{lstlisting}
\hfill
\end{minipage}

\section{Related Work}
\label{sec:related-work}

The study discussed in this paper borrowed from previous studies in the field of big data benchmarking, all of which are strictly focused on benchmarking complex big data systems and do not consider lean big data systems, with the exception of \emph{LeanBench}. This section highlights two of the more well-known big data benchmark suites, as well as LeanBench, the only previously existing benchmark which has used Unicage.

Big data benchmark suites follow and expand on the standards defined by the Transaction Processing Performance Council (TPC) and the Standard Performance Evaluation Corporation (SPEC) benchmarking consortia. The scope of big data benchmarks typically fits one of three categories~\cite{han2017benchmarking}: \emph{micro-benchmarks} are designed to evaluate individual system components through small operations; \emph{end-to-end benchmarks} are designed to evaluate whole systems through typical application scenarios; \emph{benchmark suites} are combinations of micro and end-to-end benchmarks targeted at various big data systems.

\subsection{BigDataBench}

BigDataBench~\cite{wang2014bigdatabench,gao2018bigdatabench} is a big data and artificial intelligence benchmark suite. It features micro-benchmarks for workloads comprised of single small units of computation, component benchmarks for workloads comprised of sets of small units of computation, and end-to-end benchmarks that mimic realistic application scenarios with combinations of component benchmarks. BigDataBench provides a suite of implementations of batch, query and stream processing workloads, among others, for the state-of-the-art big data systems including Spark. BDGS~\cite{ming2013bdgs}, which stands for Big Data Generation Suite, is the data generator of BigDataBench.

\subsection{BigBench}

BigBench~\cite{ghazal2013bigbench} is an end-to-end benchmark for data warehousing and query processing frameworks and mimics the use-case of a product retailer. It was first introduced in 2013 but has since been adapted into an industry standard by TPC as the TPCx-BB benchmark. PDGF~\cite{rabl2010data}, which stands for Parallel Data Generation Framework, is suitable for the generation of structured data. BigBench extends PDGF to generate semi-structured and unstructured data. In 2017, BigBench V2~\cite{ghazal2017bigbench} was proposed as a rework of the original data model and data generator to better reflect real-life scenarios.

\subsection{LeanBench}

\begin{figure}[ht]
\includegraphics[width=0.8\columnwidth]{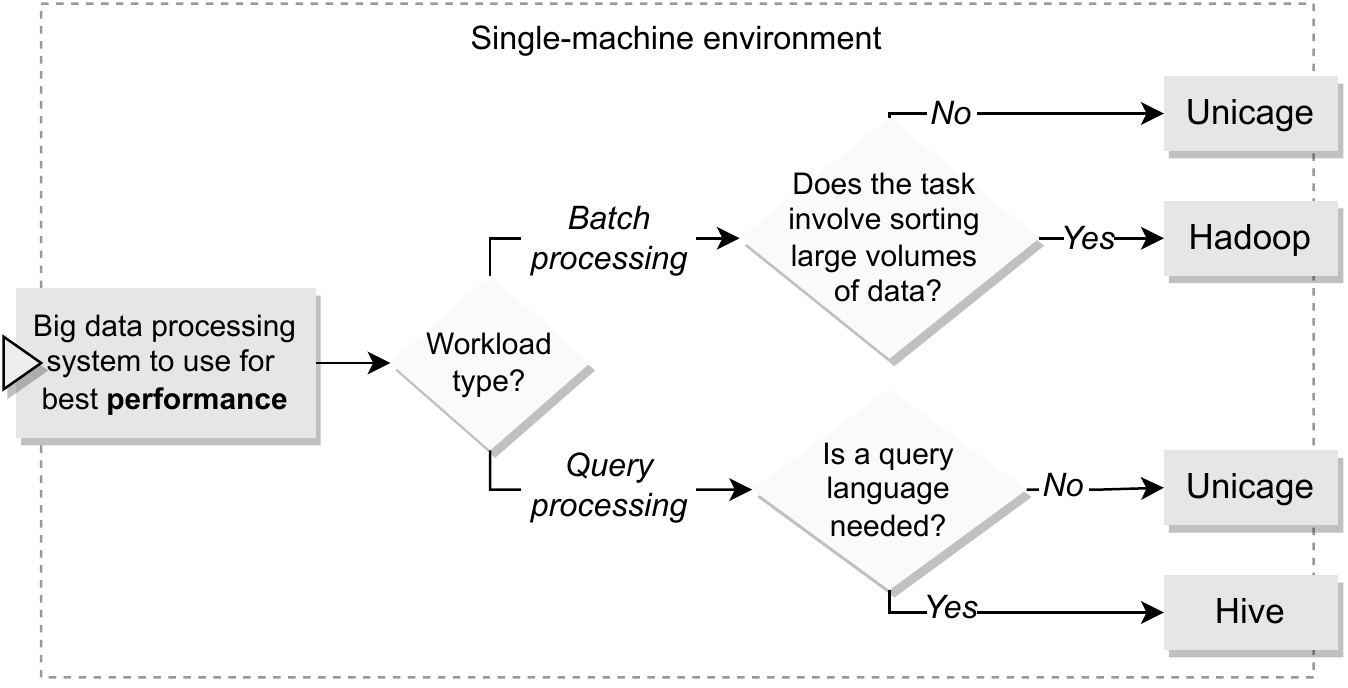}
\centering
\caption[Summary of conclusions from the LeanBench study.]{Summary of conclusions from the LeanBench study.}
\label{fig:leanbench}
\end{figure}

Moreira et al.~\cite{moreira2018leanbench} benchmarked the performance of the \emph{grep}, \emph{sort}, \emph{wordcount}, \emph{select}, \emph{join} and \emph{aggregation} workloads implemented in Unicage Tukubai, Hadoop and Hive, deployed in a single machine.
Figure~\ref{fig:leanbench} highlights the conclusions of LeanBench: in a single-machine environment, Unicage performs better than Hadoop if the task does not involve sorting large data sets (over 400\,MB), and performs better than Hive for data sets of all volumes, with the latter being the preferred system only if using a querying language is a requirement.

The main limitation of LeanBench was the deployment in a single machine which prevented tests with larger data sets, as it was not feasible to store or process terabyte-sized data sets in a single machine with commodity hardware.

\section{Experimental Set-up}
\label{sec:experimental-setup}

This section provides a detailed view of the experimental set-up, followed by an overview of the processes involved. 

\subsection{Environment}

The experimental environment was divided in three sub-clusters: the \emph{Hadoop cluster} for data processing with complex big data systems; the \emph{Unicage cluster} for data processing with lean big data systems; and the \emph{Producer cluster} for data generation and loading into the Hadoop and Unicage clusters. The Hadoop and Unicage clusters had the same specifications, as shown in Table~\ref{tab:cluster}. Every node was hosted as a Linux VSI (Virtual Server Instance) in the IBM Cloud VPC infrastructure.

\begin{table}[H]
\centering
\resizebox{\columnwidth}{!}{%
\begin{tabular}{lclccccccc}
\cline{2-10}
                                               & \multicolumn{5}{c}{Software}                                                                                                                                        & \multicolumn{4}{c}{Hardware}                                                                                \\ \hline
Hostname                                       & \multicolumn{2}{c}{Hadoop}                                             & Hive                       & Spark                      & Unicage                          & CPU              & RAM   & \begin{tabular}[c]{@{}c@{}}Network\\ bandwidth\end{tabular} & Disk               \\ \hline
\rowcolor[HTML]{FFFFFF} 
\multicolumn{10}{c}{\cellcolor[HTML]{FFFFFF}{\color[HTML]{000000} Hadoop cluster}}                                                                                                                                                                                                                                                 \\
\rowcolor[HTML]{F5F5F5} 
\textit{namenode}                              & \multicolumn{2}{c}{\cellcolor[HTML]{F5F5F5}3.3.1}                      & 3.1.2                      & 3.2.1                      & {\color[HTML]{9B9B9B} n/a}       & 8  vCPU, 2.0 GHz & 32 GB & 12 Gbps                                                     & 1024 GB, 412 Mbps  \\
\textit{datanode1...5}                         & \multicolumn{2}{c}{3.3.1}                                              & {\color[HTML]{9B9B9B} n/a} & {\color[HTML]{9B9B9B} n/a} & {\color[HTML]{9B9B9B} n/a}       & 4  vCPU, 2.0 GHz & 16 GB & 6 Gbps                                                      & 4096 GB, 1607 Mbps \\ \hline
\rowcolor[HTML]{FFFFFF} 
\multicolumn{10}{c}{\cellcolor[HTML]{FFFFFF}Unicage cluster}                                                                                                                                                                                                                                                                       \\
\rowcolor[HTML]{F5F5F5} 
\cellcolor[HTML]{F5F5F5}\textit{unicageleader} & \multicolumn{2}{c}{\cellcolor[HTML]{F5F5F5}{\color[HTML]{9B9B9B} n/a}} & {\color[HTML]{9B9B9B} n/a} & {\color[HTML]{9B9B9B} n/a} & \cellcolor[HTML]{F5F5F5}TKB, BOA & 8  vCPU, 2.0 GHz & 32 GB & 12 Gbps                                                     & 1024 GB, 412 Mbps  \\
\textit{unicageworker1...5}                    & \multicolumn{2}{c}{{\color[HTML]{9B9B9B} n/a}}                         & {\color[HTML]{9B9B9B} n/a} & {\color[HTML]{9B9B9B} n/a} & TKB, BOA                     & 4  vCPU, 2.0 GHz & 16 GB & 6 Gbps                                                      & 4096 GB, 1607 Mbps \\ \hline
\rowcolor[HTML]{FFFFFF} 
\multicolumn{10}{c}{\cellcolor[HTML]{FFFFFF}Producer cluster}                                                                                                                                                                                                                                                                      \\
\rowcolor[HTML]{F5F5F5} 
\cellcolor[HTML]{F5F5F5}\textit{producer1...4} & \multicolumn{2}{c}{\cellcolor[HTML]{F5F5F5}3.3.1}                      & {\color[HTML]{9B9B9B} n/a} & {\color[HTML]{9B9B9B} n/a} & TKB, BOA                     & 8  vCPU, 2.0 GHz & 32 GB & \cellcolor[HTML]{F5F5F5}12 Gbps                             & 4096 GB, 1607 Mbps \\ \hline
\end{tabular} %
}
\caption{Specifications of the experimental cluster environment.}
\label{tab:cluster}
\end{table}

The nodes in the Producer cluster were configured as HDFS clients to allow data loading into the Hadoop cluster; and the relevant Unicage commands were also installed to allow data loading into the Unicage cluster.

The benchmarks did not contemplate fault tolerance, so replication was turned off in the Hadoop cluster, by setting the \texttt{dfs.replication} property to 0 in the configuration file of HDFS. Some of the largest input data set volumes used in the benchmarks exceeded the maximum number of file blocks per file allowed in HDFS, of 10000, by a small margin, so to load these files into the Hadoop cluster, this limit was lifted by setting the \texttt{dfs.namenode.fs-limits.max-blocks-per-file} property to a sufficiently large number in the configuration file of HDFS. Spark was configured to take advantage of the four CPU cores of the five DataNodes in the Hadoop cluster, by setting the \texttt{spark.executor.instances} and \texttt{spark.executor.cores} configuration properties to 5 and 4, respectively.
The caching level of Spark was left with the default configuration: when it runs out of memory, it uses the lineage information of RDDs to recompute them on demand, allowing the workloads to proceed with reduced performance. Spark was also configured to use the Metastore of Hive for access to the tables and schema of the structured data sets. 

\subsection{Experimental Overview}

\begin{figure}[ht]
\includegraphics[width=0.8\columnwidth]{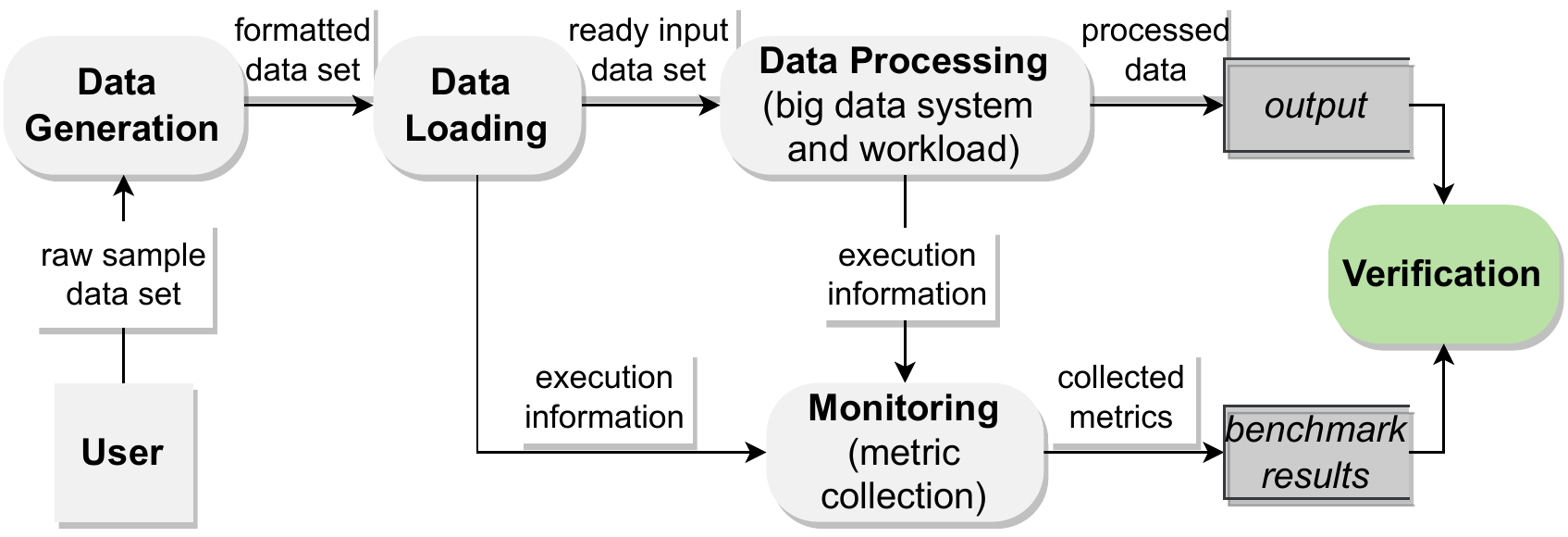}
\centering
\caption[Data-flow diagram of a big data benchmarking experiment.]{Data-flow diagram of a big data benchmarking experiment.}
\label{fig:data-flow-experiment}
\end{figure}

Two sets of benchmarks were performed: a set of batch processing workloads over unstructured data sets, and a set of query processing workloads over structured data sets. Each benchmark had a similar set of steps, as represented by the data-flow diagram in Figure~\ref{fig:data-flow-experiment}.

\subsubsection{Data Generation}

Data was generated in the Producer cluster. 
The unstructured data sets simulating \emph{Wikipedia text entries} were generated with BDGS and were used for the batch processing benchmarks. 
The structured data sets simulating \emph{e-commerce item/order tables} were generated with the demo version of PDGF\footnote{The demo version of PDGF allows parallel generation of structured data in a single-machine, while the full version allows distributed data generation in a cluster. Unfortunately, the full version was not available, so we purposefully extended the demo version to allow distributed data generation.} and were used for the query processing benchmarks. The planned volumes of the input data sets ranged from 64\,GB to 8192\,GB, but the generated volumes had some variability in size.

The structured e-commerce item/order tables data sets featured two tables, whose sizes are listed in Table~\ref{tab:query-gen-tables}.

\begin{table}[H]
\centering
\begin{tabular}{lcccccccc}
\cline{2-9}
                                          & \multicolumn{8}{c}{Real volumes (GB)}                                                                                                                                                                                                                                                                                                                                                                                                                                                                                      \\ \hline
Data set                                  & 68                                                                                & 139                                                       & 281                                                       & 565                                                       & 1145                                                      & 2320                                                      & 4672                                                      & 9417                                                      \\ \hline
\rowcolor[HTML]{F5F5F5} 
\cellcolor[HTML]{F5F5F5}\emph{Item} table & \cellcolor[HTML]{F5F5F5}\begin{tabular}[c]{@{}c@{}}43\end{tabular} & \begin{tabular}[c]{@{}c@{}}86\end{tabular} & \begin{tabular}[c]{@{}c@{}}173\end{tabular} & \begin{tabular}[c]{@{}c@{}}347\end{tabular} & \begin{tabular}[c]{@{}c@{}}698\end{tabular} & \begin{tabular}[c]{@{}c@{}}1407\end{tabular} & \begin{tabular}[c]{@{}c@{}}2823\end{tabular} & \begin{tabular}[c]{@{}c@{}}5672\end{tabular} \\
\emph{Order} table                         & \begin{tabular}[c]{@{}c@{}}26\end{tabular} & \begin{tabular}[c]{@{}c@{}}53\end{tabular} & \begin{tabular}[c]{@{}c@{}}108\end{tabular} & \begin{tabular}[c]{@{}c@{}}218\end{tabular} & \begin{tabular}[c]{@{}c@{}}447\end{tabular} & \begin{tabular}[c]{@{}c@{}}914\end{tabular} & \begin{tabular}[c]{@{}c@{}}1849\end{tabular} & \begin{tabular}[c]{@{}c@{}}3745\end{tabular}                         \\ \hline
\end{tabular}%
\caption{Data generation: the real generated table volumes of the e-commerce tables structured data sets.}
\label{tab:query-gen-tables}
\end{table}

The two data sets also had distinct layouts in the file system. The Wikipedia text entries data sets were divided in files with a fixed average volume of about 500\,MB each. A sufficient increase in the volume of these data sets resulted in an increased number of files. The e-commerce item/order tables data sets were divided in a fixed number of files. Any increase in the volume of these data sets resulted in increased volumes for each file. Table~\ref{tab:number_and_volume_files} highlights this dichotomy for the largest tested volumes.

\begin{table}[H]
\centering
\begin{tabular}{lcrr}
\hline
{\color[HTML]{000000} Data set}                & {\color[HTML]{000000} Real volume} & {\color[HTML]{000000} Number of files} & {\color[HTML]{000000} Volume per file} \\ \hline
\rowcolor[HTML]{F5F5F5} 
{\color[HTML]{000000} Wikipedia text entries}  & {\color[HTML]{000000} 8019 GB}     & {\color[HTML]{000000} 16380 files}     & {\color[HTML]{000000} 0.490 GB}        \\
{\color[HTML]{000000} E-commerce item tables}  & {\color[HTML]{000000} 5672 GB}     & {\color[HTML]{000000} 4 files}         & {\color[HTML]{000000} 1418 GB}         \\
\rowcolor[HTML]{F5F5F5} 
{\color[HTML]{000000} E-commerce order tables} & {\color[HTML]{000000} 3745 GB}     & {\color[HTML]{000000} 4 files}         & {\color[HTML]{000000} 936 GB}          \\ \hline
\end{tabular}%
\caption{Number of files and estimated volume per file of the largest tested input data sets.}
\label{tab:number_and_volume_files}
\end{table}

\subsubsection{Data Loading}

The input data sets were loaded from the Producer cluster into the the Hadoop cluster with the \texttt{put} operation of HDFS, and loaded into the Unicage cluster with the \texttt{distr-distr}\footnotemark~BOA command.
\footnotetext{\texttt{distr-distr} splits a file in a number of chunks equal to the number of Unicage worker nodes, and places one chunk in each worker, in a specified directory.}

After the data loading process, both the Hadoop and Unicage clusters had the same data sets loaded and ready for the data processing benchmarks.

\subsubsection{Data Processing}

\emph{Batch processing} workloads handle unstructured data sets.
\emph{Query processing} workloads handle structured data sets through the use of a querying language.
Both are performed on fully available bounded data sets.
%
Table~\ref{tab:benchmarked-workloads} lists the benchmarked combinations of data processing workload and big data system.

\begin{table}[H]
\centering
\resizebox{\columnwidth}{!}{%
\begin{tabular}{llllll}
\cline{2-5}
                     & \multicolumn{4}{c}{Big data systems}                                    &                                                                                                                        \\ \hline
Workload             & Hadoop                    & Hive                      & Spark & Unicage & Description                                                                                                            \\ \hline
\rowcolor[HTML]{FFFFFF} 
\multicolumn{6}{c}{\cellcolor[HTML]{FFFFFF}Batch processing - Wikipedia text entries data set}                                                                                                                          \\
\rowcolor[HTML]{F5F5F5} 
\textit{grep}        & Yes                       & {\color[HTML]{9B9B9B} No} & Yes   & Yes     & \begin{tabular}[c]{@{}l@{}}Count the number of occurrences of a substring in the text data set.\end{tabular}        \\
\textit{sort}        & Yes                       & {\color[HTML]{9B9B9B} No} & Yes   & Yes     & \begin{tabular}[c]{@{}l@{}}Sort a single-column rendition of the text data set, with one word per row.\end{tabular} \\
\rowcolor[HTML]{F5F5F5} 
\textit{wordcount}   & Yes                       & {\color[HTML]{9B9B9B} No} & Yes   & Yes     & \begin{tabular}[c]{@{}l@{}}Count the number of occurrences of each unique word in the text data set.\end{tabular}   \\ \hline
\rowcolor[HTML]{FFFFFF} 
\multicolumn{6}{c}{\cellcolor[HTML]{FFFFFF}Query processing - e-commerce item/order tables data set}                                                                                                                    \\
\rowcolor[HTML]{F5F5F5} 
\textit{select}      & {\color[HTML]{9B9B9B} No} & Yes                       & Yes   & Yes     & \begin{tabular}[c]{@{}l@{}}Select rows of the \emph{item} table whose target column exceeds a given value.\end{tabular}    \\
\textit{join}        & {\color[HTML]{9B9B9B} No} & Yes                       & Yes   & Yes     & \begin{tabular}[c]{@{}l@{}}Combine rows of the \emph{item} and \emph{order} tables based on a related column.\end{tabular}         \\
\rowcolor[HTML]{F5F5F5} 
\textit{aggregation} & {\color[HTML]{9B9B9B} No} & Yes                       & Yes   & Yes     & \begin{tabular}[c]{@{}l@{}}Sums the values of column A of the \emph{item} table, grouped by column B.\end{tabular}        \\ \hline
\end{tabular}
}
\caption{Benchmarked workloads and big data systems.}
\label{tab:benchmarked-workloads}
\end{table}

The benchmarked workloads are commonly seen micro-benchmarks in batch and query processing big data benchmarks:
\emph{grep} and \emph{select} are search workloads; \emph{wordcount} and \emph{aggregation} are grouping workloads; \emph{sort} and \emph{join} are I/O-intensive workloads.

\subsubsection{Monitoring}

The data loading and data processing workloads were monitored through the collection of a variety of metrics. \emph{Execution time} was the main performance measure, and \emph{loading/processing rate} was calculated to capture performance fluctuations between benchmarks with different input data set volumes. Additionally, Netdata\footnote{Netdata is a tool for cluster observability. https://www.netdata.cloud/} was used to collect the system resource usage metrics that allowed a better understanding of the resource usage patterns and aided in the configuration and optimization of the big data systems and workloads.

\subsubsection{Verification}
\label{verification}

In practice, it was not feasible to verify the absolute correctness of the outputs for the benchmarked data set volumes, as these were very large. 
The verification process merely assessed what output was more likely to be correct, based on the mutual agreement between the systems.
The output of each combination of data processing workload and big data system was verified alongside the outputs of peer combinations.
For any given workload, to efficiently assess the agreement between the tested big data systems, the output of each system was used to compute its MD5 hash and the resulting hashes were compared.
If the hashes of the outputs of Unicage, Hadoop and Spark were the same, there was \emph{agreement}, and the output was assumed to be \emph{correct}.

\subsubsection{Validation}

Every benchmark was repeated thrice, and both the sampled execution times and loading/processing rates were statistically validated through the estimation of the population mean and confidence intervals, as described by Montgomery~\cite{montgomery2010applied}.

\section{Experimental Results}
\label{sec:experimental-results}

This section lists and discusses the results produced by the benchmarks.

\subsection{Batch Processing Benchmarks}

The 
Wikipedia text entries 
were used as input data sets for the batch processing workloads in this benchmark set.

\subsubsection{Data Loading - Unstructured Data Set}

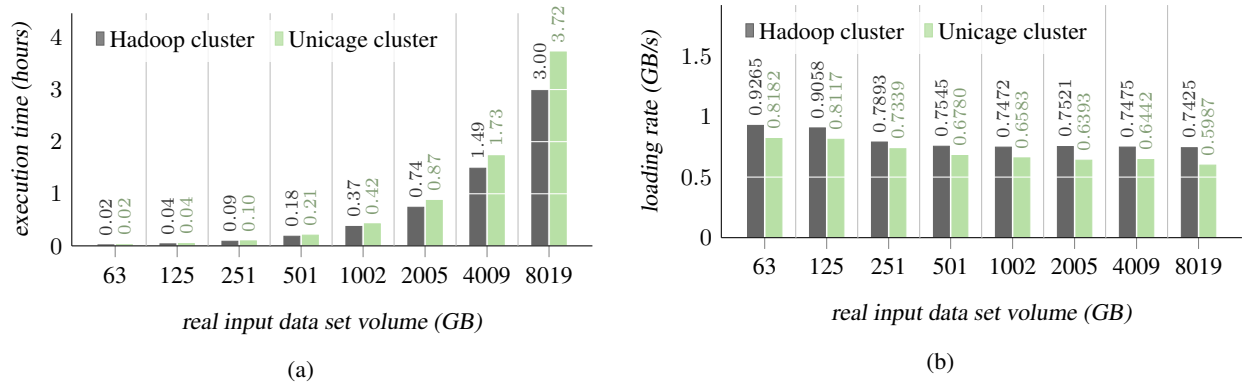
\begin{figure}[h]
\makebox[\textwidth][c]{
\begin{subfigure}{.515\textwidth}
\begin{tikzpicture}
\begin{axis}[
    ybar = 1pt,
    bar width=6pt,
	xtick = {0,1,2,3,4,5,6,7},
	xticklabels = {63,125,251,501,1002,2005,4009,8019},
    x tick label style={
        /pgf/number format/1000 sep=, font=\small
    },
    y tick label style={
        font=\small
    },
    enlarge x limits = 0.1,
	width = \columnwidth,
	height = 0.55\columnwidth,
	xlabel = real input data set volume (GB),
	ylabel = execution time (hours),
	legend cell align = {left},
	legend pos = north west,
    xmin = 0,
    xmode = normal,
	ymajorgrids=true,
	major grid style = {lightgray},
	minor grid style = {lightgray!25},
	compat=newest,
	nodes near coords,
	nodes near coords style={font=\scriptsize, rotate=90, anchor=west, /pgf/number format/.cd, fixed, fixed zerofill, precision=2, /tikz/.cd},
	enlarge y limits={upper, value=0.2},
	axis x line*=bottom,
	axis y line*=left,
	ymin=0,
	ytick style={draw=none},
	cycle list={
		{fill=black!60,draw=black!60,text=black!80},
		{fill=bargreen!100,draw=bargreen!100,text=bargreen!70!black}
	},
	legend image code/.code={
        \draw [#1] (0cm,-0.1cm) rectangle (0.2cm,0.25cm); },
	axis on top,
	major grid style=white,
	ymajorgrids,
	extra x ticks={0.5,1.5,2.5,3.5,4.5,5.5,6.5},
    extra x tick labels={},
    extra x tick style={
        grid=major,
        major tick length=0pt,
		major grid style = {lightgray},
    },
        legend image post style={scale=0.5},
	legend style={draw=none,fill opacity=0.8,text opacity=1,/tikz/every even column/.append style={column sep=0.2cm}, legend columns=-1, font=\footnotesize}
]
\addplot table [x expr=\coordindex, y expr=\thisrow{hadoop}/3600, col sep=comma] {CSVs/batch-data-loading.csv};
\addplot table [x expr=\coordindex, y expr=\thisrow{unicage}/3600, col sep=comma] {CSVs/batch-data-loading.csv};
\legend{Hadoop cluster,Unicage cluster}
\end{axis}
\end{tikzpicture}
\centering
\caption{}
\label{fig:batch-data-loading}
\end{subfigure}%
\begin{subfigure}{.515\textwidth}
\begin{tikzpicture}
\begin{axis}[
    ybar = 1pt,
    bar width=6pt,
	xtick = {0,1,2,3,4,5,6,7},
	xticklabels = {63,125,251,501,1002,2005,4009,8019},
    x tick label style={
        /pgf/number format/1000 sep=, font=\small
    },
    y tick label style={
        font=\small
    },
    enlarge x limits = 0.1,
	width = \columnwidth,
	height = 0.55\columnwidth,
	xlabel = real input data set volume (GB),
	ylabel = loading rate (GB/s),
	legend cell align = {left},
	legend pos = north west,
    xmin = 0,
    xmode = normal,
	ymajorgrids=true,
	major grid style = {lightgray},
	minor grid style = {lightgray!25},
	compat=newest,
	nodes near coords,
	nodes near coords style={font=\scriptsize, rotate=90, anchor=west, /pgf/number format/.cd, fixed, fixed zerofill, precision=4, /tikz/.cd},
	enlarge y limits={upper, value=0.2},
	axis x line*=bottom,
	axis y line*=left,
	ymin=0,
	ymax=1.6,
	ytick style={draw=none},
	cycle list={
		{fill=black!60,draw=black!60,text=black!80},
		{fill=bargreen!100,draw=bargreen!100,text=bargreen!70!black}
	},
	legend image code/.code={
        \draw [#1] (0cm,-0.1cm) rectangle (0.2cm,0.25cm); },
	axis on top,
	major grid style=white,
	ymajorgrids,
	extra x ticks={0.5,1.5,2.5,3.5,4.5,5.5,6.5},
    extra x tick labels={},
    extra x tick style={
        grid=major,
        major tick length=0pt,
		major grid style = {lightgray},
    },
        legend image post style={scale=0.5},
	legend style={draw=none,fill opacity=0.8,text opacity=1,/tikz/every even column/.append style={column sep=0.2cm}, legend columns=-1, font=\footnotesize}
]
\addplot table [x expr=\coordindex, y expr=\thisrow{realsize}/\thisrow{hadoop}, col sep=comma] {CSVs/batch-data-loading.csv};
\addplot table [x expr=\coordindex, y expr=\thisrow{realsize}/\thisrow{unicage}, col sep=comma] {CSVs/batch-data-loading.csv};
\legend{Hadoop cluster,Unicage cluster}
\end{axis}
\end{tikzpicture}
\centering
\caption{}
\label{fig:batch-data-loading-dps}
\end{subfigure}
}
\caption{Data loading execution times (a) and rates (b) for the Wikipedia text entries unstructured data set.}
\end{figure}

Figures~\ref{fig:batch-data-loading} and~\ref{fig:batch-data-loading-dps} show the execution times and loading rates, respectively, of loading the 
Wikipedia text entries 
into both the Hadoop and Unicage clusters, for the tested input data set volumes. 

These two types of bar charts are recurrent for the upcoming results.
Bar charts pertaining to the \emph{execution times} highlight the performance differences between the big data systems, for the set of tested input data set volumes. In these charts: lower time bars mean better performance. The vertical axis shows the execution time, in hours, and the horizontal axis shows the set of tested input data set volumes, in gigabytes.

The bar charts pertaining to the \emph{loading} or \emph{processing rates} provide an understanding of the performance fluctuations of each big data system across input data set volumes, which are, otherwise, not perceptible.
In these charts: higher rate bars mean better performance. The vertical axis shows the loading rate (for data loading workloads) or processing rate (for data processing workloads) in gigabytes per second, and the horizontal axis shows the set of tested input data set volumes, in gigabytes.

In both types of chart, for easy reading, grey vertical lines mark the divisions between data set volumes. 

Loading the unstructured Wikipedia text entries data sets into the Unicage cluster was slightly slower than loading the same data sets into the Hadoop cluster, across all tested input data set volumes.

\subsubsection{Grep}

\begin{figure}[h]
\makebox[\textwidth][c]{
\begin{subfigure}{.515\textwidth}
\begin{tikzpicture}
\begin{axis}[
    ybar = 1pt,
    bar width=6pt,
	xtick = {0,1,2,3,4,5,6,7},
	xticklabels = {63,125,251,501,1002,2005,4009,8019},
    x tick label style={
        /pgf/number format/1000 sep=, font=\small
    },
    y tick label style={
        font=\small
    },
    enlarge x limits = 0.1,
	width = \columnwidth,
	height = 0.55\columnwidth,
	xlabel = real input data set volume (GB),
	ylabel = execution time (hours),
	legend cell align = {left},
	legend pos = north west,
    xmin = 0,
    xmode = normal,
	ymajorgrids=true,
	major grid style = {lightgray},
	minor grid style = {lightgray!25},
	compat=newest,
	nodes near coords,
	nodes near coords style={font=\scriptsize, rotate=90, anchor=west, /pgf/number format/.cd, fixed, fixed zerofill, precision=2, /tikz/.cd},
	enlarge y limits={upper, value=0.2},
	axis x line*=bottom,
	axis y line*=left,
	ymin=0,
	ytick style={draw=none},
	cycle list={
		{fill=black!60,draw=black!60,text=black!80},
		{fill=sparkblue,draw=sparkblue,text=sparkblue!70!black},
		{fill=bargreen!100,draw=bargreen!100,text=bargreen!70!black}
	},
	legend image code/.code={
        \draw [#1] (0cm,-0.1cm) rectangle (0.2cm,0.25cm); },
	axis on top,
	major grid style=white,
	ymajorgrids,
	extra x ticks={0.5,1.5,2.5,3.5,4.5,5.5,6.5},
    extra x tick labels={},
    extra x tick style={
        grid=major,
        major tick length=0pt,
		major grid style = {lightgray},
    },
        legend image post style={scale=0.5},
	legend style={draw=none,fill opacity=0.8,text opacity=1,/tikz/every even column/.append style={column sep=0.2cm}, legend columns=-1, font=\footnotesize}
]
\addplot table [x expr=\coordindex, y expr=\thisrow{hadoop}/3600, col sep=comma] {CSVs/grep-execution-time.csv};
\addplot table [x expr=\coordindex, y expr=\thisrow{spark}/3600, col sep=comma] {CSVs/grep-execution-time.csv};
\addplot table [x expr=\coordindex, y expr=\thisrow{unicage}/3600, col sep=comma] {CSVs/grep-execution-time.csv};
\legend{Hadoop MR,Spark,Unicage}
\end{axis}
\end{tikzpicture}
\centering
\caption{}
\label{fig:grep-execution-time}
\end{subfigure}%
\begin{subfigure}{.515\textwidth}
\begin{tikzpicture}
\begin{axis}[
    ybar = 1pt,
    bar width=6pt,
	xtick = {0,1,2,3,4,5,6,7},
	xticklabels = {63,125,251,501,1002,2005,4009,8019},
    x tick label style={
        /pgf/number format/1000 sep=, font=\small
    },
    y tick label style={
        font=\small
    },
    enlarge x limits = 0.1,
	width = \columnwidth,
	height = 0.55\columnwidth,
	xlabel = real input data set volume (GB),
	ylabel = processing rate (GB/s),
	legend cell align = {left},
	legend pos = north west,
    xmin = 0,
    xmode = normal,
	ymajorgrids=true,
	major grid style = {lightgray},
	minor grid style = {lightgray!25},
	compat=newest,
	nodes near coords,
	nodes near coords style={font=\scriptsize, rotate=90, anchor=west, /pgf/number format/.cd, fixed, fixed zerofill, precision=4, /tikz/.cd},
	enlarge y limits={upper, value=0.2},
	axis x line*=bottom,
	axis y line*=left,
	ymin=0,
	ymax=1.60,
	ytick style={draw=none},
	cycle list={
		{fill=black!60,draw=black!60,text=black!80},
		{fill=sparkblue,draw=sparkblue,text=sparkblue!70!black},
		{fill=bargreen!100,draw=bargreen!100,text=bargreen!70!black}
	},
	legend image code/.code={
        \draw [#1] (0cm,-0.1cm) rectangle (0.2cm,0.25cm); },
	axis on top,
	major grid style=white,
	ymajorgrids,
	extra x ticks={0.5,1.5,2.5,3.5,4.5,5.5,6.5},
    extra x tick labels={},
    extra x tick style={
        grid=major,
        major tick length=0pt,
		major grid style = {lightgray},
    },
        legend image post style={scale=0.5},
	legend style={draw=none,fill opacity=0.8,text opacity=1,/tikz/every even column/.append style={column sep=0.2cm}, legend columns=-1, font=\footnotesize}
]
\addplot table [x expr=\coordindex, y expr=\thisrow{realsize}/\thisrow{hadoop}, col sep=comma] {CSVs/grep-execution-time.csv};
\addplot table [x expr=\coordindex, y expr=\thisrow{realsize}/\thisrow{spark}, col sep=comma] {CSVs/grep-execution-time.csv};
\addplot table [x expr=\coordindex, y expr=\thisrow{realsize}/\thisrow{unicage}, col sep=comma] {CSVs/grep-execution-time.csv};
\legend{Hadoop MR,Spark,Unicage}
\end{axis}
\end{tikzpicture}
\centering
\caption{}
\label{fig:grep-dps}
\end{subfigure}
}
\caption{Data processing execution times (a) and rates (b) for \textit{grep}.}
\end{figure}
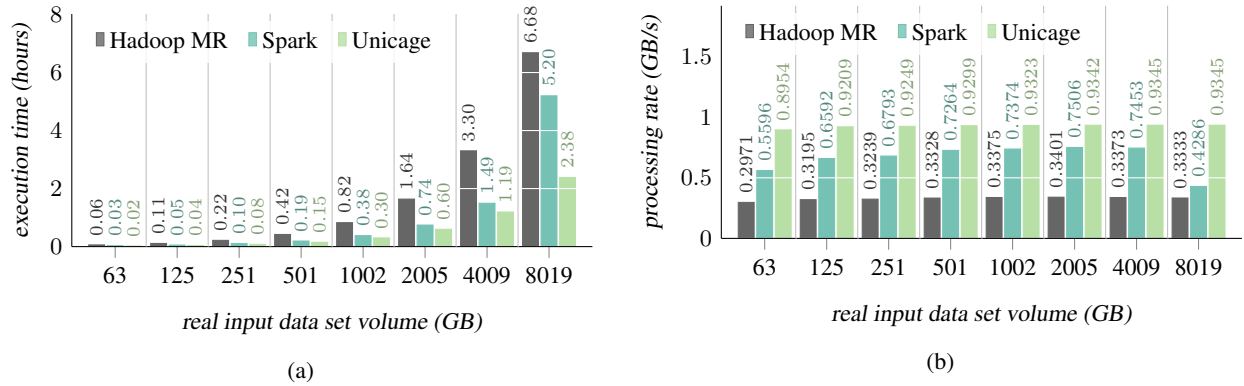

Figures~\ref{fig:grep-execution-time} and~\ref{fig:grep-dps} show the execution times and processing rates, respectively, of each big data system when performing the \emph{grep} workload for the tested input data set volumes.

The performance of the \emph{grep} implementation with Unicage was superior to those of Hadoop and Spark, across all tested input data set volumes. 

A recurrent pattern in the processing rate charts is the reduced processing rate of Spark for smaller data sets. This is caused by the warm-up period of Spark, when the Driver process is started, along with the SparkContext and when the resource negotiation with YARN takes place. This is more perceivable in workloads that take very little time to complete, such is the case of the \emph{grep} workload for the 63\,GB and 125\,GB input data sets.

The processing rate of Spark when performing the \emph{grep} workload on the 8019\,GB input data set is visibly reduced. This was caused by the performance degradation of Spark when it starts recomputing RDD partitions on demand. A similar performance degradation was observed in the \emph{select}, \emph{join} and \emph{aggregation} workloads. 

\subsubsection{Sort}

\begin{figure}[h]
\makebox[\textwidth][c]{
\begin{subfigure}{.515\textwidth}
\begin{tikzpicture}
\begin{axis}[
    ybar = 1pt,
    bar width=6pt,
	xtick = {0,1,2,3},
	xticklabels = {63,125,251,501},
    x tick label style={
        /pgf/number format/1000 sep=, font=\small
    },
    y tick label style={
        font=\small
    },
    enlarge x limits = 0.1,
	width = \columnwidth,
	height = 0.55\columnwidth,
	xlabel = real input data set volume (GB),
	ylabel = execution time (hours),
	legend cell align = {left},
	legend pos = north west,
    xmin = 0,
    xmode = normal,
	ymajorgrids=true,
	major grid style = {lightgray},
	minor grid style = {lightgray!25},
	compat=newest,
	nodes near coords,
	nodes near coords style={font=\scriptsize, rotate=90, anchor=west, /pgf/number format/.cd, fixed, fixed zerofill, precision=2, /tikz/.cd},
	enlarge y limits={upper, value=0.2},
	axis x line*=bottom,
	axis y line*=left,
	ymin=0,
	ytick style={draw=none},
	cycle list={
		{fill=black!60,draw=black!60,text=black!80},
		{fill=sparkblue,draw=sparkblue,text=sparkblue!70!black},
		{fill=bargreen!100,draw=bargreen!100,text=bargreen!70!black}
	},
	legend image code/.code={
        \draw [#1] (0cm,-0.1cm) rectangle (0.2cm,0.25cm); },
	axis on top,
	major grid style=white,
	ymajorgrids,
	extra x ticks={0.5,1.5,2.5,3.5,4.5,5.5,6.5},
    extra x tick labels={},
    extra x tick style={
        grid=major,
        major tick length=0pt,
		major grid style = {lightgray},
    },
        legend image post style={scale=0.5},
	legend style={draw=none,fill opacity=0.8,text opacity=1,/tikz/every even column/.append style={column sep=0.2cm}, legend columns=2, transpose legend, font=\footnotesize}
]
\addplot table [x expr=\coordindex, y expr=\thisrow{hadoop}/3600, col sep=comma] {CSVs/sort-execution-time.csv};
\addplot table [x expr=\coordindex, y expr=\thisrow{spark}/3600, col sep=comma] {CSVs/sort-execution-time.csv};
\addplot table [x expr=\coordindex, y expr=\thisrow{unicage}/3600, col sep=comma] {CSVs/sort-execution-time.csv};
\legend{Hadoop MR,Spark,Unicage}
\end{axis}
\end{tikzpicture}
\centering
\caption{}
\label{fig:sort-execution-time}
\end{subfigure}
\begin{subfigure}{.515\textwidth}
\begin{tikzpicture}
\begin{axis}[
    ybar = 1pt,
    bar width=6pt,
	xtick = {0,1,2,3},
	xticklabels = {63,125,251,501},
    x tick label style={
        /pgf/number format/1000 sep=, font=\small
    },
    y tick label style={
        font=\small
    },
	ytick = {0,0.0200,0.0400,0.0600},
    enlarge x limits = 0.1,
	width = \columnwidth,
	height = 0.55\columnwidth,
	xlabel = real input data set volume (GB),
	ylabel = processing rate (GB/s),
	legend cell align = {left},
	legend pos = north west,
    xmin = 0,
    xmode = normal,
	ymajorgrids=true,
	major grid style = {lightgray},
	minor grid style = {lightgray!25},
	compat=newest,
	nodes near coords,
	nodes near coords style={font=\scriptsize, rotate=90, anchor=west, /pgf/number format/.cd, fixed, fixed zerofill, precision=4, /tikz/.cd},
	enlarge y limits={upper, value=0.2},
	axis x line*=bottom,
	axis y line*=left,
	ymin=0,
	ymax=0.07,
	ytick style={draw=none},
	cycle list={
		{fill=black!60,draw=black!60,text=black!80},
		{fill=sparkblue,draw=sparkblue,text=sparkblue!70!black},
		{fill=bargreen!100,draw=bargreen!100,text=bargreen!70!black}
	},
	legend image code/.code={
        \draw [#1] (0cm,-0.1cm) rectangle (0.2cm,0.25cm); },
	axis on top,
	major grid style=white,
	ymajorgrids,
	extra x ticks={0.5,1.5,2.5,3.5,4.5,5.5,6.5},
    extra x tick labels={},
    extra x tick style={
        grid=major,
        major tick length=0pt,
		major grid style = {lightgray},
    },
        legend image post style={scale=0.5},
	legend style={draw=none,fill opacity=0.8,text opacity=1,/tikz/every even column/.append style={column sep=0.2cm}, legend columns=-1, font=\footnotesize}
]
\addplot table [x expr=\coordindex, y expr=\thisrow{realsize}/\thisrow{hadoop}, col sep=comma] {CSVs/sort-execution-time.csv};
\addplot table [x expr=\coordindex, y expr=\thisrow{realsize}/\thisrow{spark}, col sep=comma] {CSVs/sort-execution-time.csv};
\addplot table [x expr=\coordindex, y expr=\thisrow{realsize}/\thisrow{unicage}, col sep=comma] {CSVs/sort-execution-time.csv};
\legend{Hadoop MR,Spark,Unicage}
\end{axis}
\end{tikzpicture}
\centering
\caption{}
\label{fig:sort-dps}
\end{subfigure}
}
\caption{Data processing execution times (a) and rates (b) for \textit{sort}.}
\end{figure}
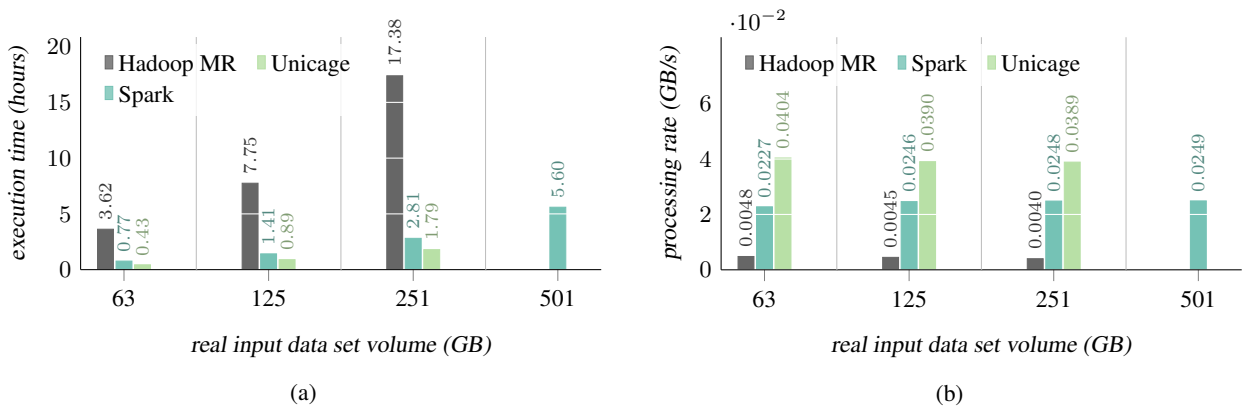

Figures~\ref{fig:sort-execution-time} and~\ref{fig:sort-dps} show the execution times and processing rates, respectively, of each big data system under test, when performing the \emph{sort} workload for the tested input data set volumes.

The performance of the \emph{sort} implementation with Unicage was superior to those of Hadoop and Spark for the 63\,GB, 125\,GB and 251\,GB input data sets, but produced arbitrarily incorrect outputs for the 501\,GB and larger input data sets.
Ideally, a big data system should either produce a correct output or finish with an identified fault, but the aforementioned results show that Unicage failed to achieve either outcome.

The output of the \emph{sort} implementation with Spark for the 501\,GB input data set was verified through the comparison with the output of a single run of the \emph{sort} implementation with Hadoop, which took approximately 38 hours to complete, and thus was not validated statistically. The \emph{sort} workload was not attempted for larger data sets as the experimental setup did not allow the efficient verification of the outputs of Spark.

\subsubsection{Wordcount}

\begin{figure}[h]
\makebox[\textwidth][c]{
\begin{subfigure}{.515\textwidth}
\begin{tikzpicture}
\begin{axis}[
    ybar = 1pt,
    bar width=6pt,
	xtick = {0,1,2,3,4,5,6,7},
	xticklabels = {63,125,251,501,1002,2005,4009,8019},
    x tick label style={
        /pgf/number format/1000 sep=, font=\small
    },
    y tick label style={
        font=\small
    },
    enlarge x limits = 0.1,
	width = \columnwidth,
	height = 0.55\columnwidth,
	xlabel = real input data set volume (GB),
	ylabel = execution time (hours),
	legend cell align = {left},
	legend pos = north west,
    xmin = 0,
    xmode = normal,
	ymajorgrids=true,
	major grid style = {lightgray},
	minor grid style = {lightgray!25},
	compat=newest,
	nodes near coords,
	nodes near coords style={font=\scriptsize, rotate=90, anchor=west, /pgf/number format/.cd, fixed, fixed zerofill, precision=2, /tikz/.cd},
	enlarge y limits={upper, value=0.2},
	axis x line*=bottom,
	axis y line*=left,
	ymin=0,
	ytick style={draw=none},
	cycle list={
		{fill=black!60,draw=black!60,text=black!80},
		{fill=sparkblue,draw=sparkblue,text=sparkblue!70!black},
		{fill=bargreen!100,draw=bargreen!100,text=bargreen!70!black}
	},
	legend image code/.code={
        \draw [#1] (0cm,-0.1cm) rectangle (0.2cm,0.25cm); },
	axis on top,
	major grid style=white,
	ymajorgrids,
	extra x ticks={0.5,1.5,2.5,3.5,4.5,5.5,6.5},
    extra x tick labels={},
    extra x tick style={
        grid=major,
        major tick length=0pt,
		major grid style = {lightgray},
    },
        legend image post style={scale=0.5},
	legend style={draw=none,fill opacity=0.8,text opacity=1,/tikz/every even column/.append style={column sep=0.2cm}, legend columns=-1, font=\footnotesize}
]
\addplot table [x expr=\coordindex, y expr=\thisrow{hadoop}/3600, col sep=comma] {CSVs/wordcount-execution-time.csv};
\addplot table [x expr=\coordindex, y expr=\thisrow{spark}/3600, col sep=comma] {CSVs/wordcount-execution-time.csv};
\addplot table [x expr=\coordindex, y expr=\thisrow{unicage}/3600, col sep=comma] {CSVs/wordcount-execution-time.csv};
\legend{Hadoop MR,Spark,Unicage}
\end{axis}
\end{tikzpicture}
\centering
\caption{}
\label{fig:wordcount-execution-time}
\end{subfigure}
\begin{subfigure}{.515\textwidth}
\begin{tikzpicture}
\begin{axis}[
    ybar = 1pt,
    bar width=6pt,
	xtick = {0,1,2,3,4,5,6,7},
	xticklabels = {63,125,251,501,1002,2005,4009,8019},
    x tick label style={
        /pgf/number format/1000 sep=, font=\small
    },
    y tick label style={
        font=\small
    },
    enlarge x limits = 0.1,
	width = \columnwidth,
	height = 0.55\columnwidth,
	xlabel = real input data set volume (GB),
	ylabel = processing rate (GB/s),
	legend cell align = {left},
	legend pos = north west,
    xmin = 0,
    xmode = normal,
	ymajorgrids=true,
	major grid style = {lightgray},
	minor grid style = {lightgray!25},
	compat=newest,
	nodes near coords,
	nodes near coords style={font=\scriptsize, rotate=90, anchor=west, /pgf/number format/.cd, fixed, fixed zerofill, precision=4, /tikz/.cd},
	enlarge y limits={upper, value=0.2},
	axis x line*=bottom,
	axis y line*=left,
	ymin=0,
	ymax=0.6,
	ytick style={draw=none},
	cycle list={
		{fill=black!60,draw=black!60,text=black!80},
		{fill=sparkblue,draw=sparkblue,text=sparkblue!70!black},
		{fill=bargreen!100,draw=bargreen!100,text=bargreen!70!black}
	},
	legend image code/.code={
        \draw [#1] (0cm,-0.1cm) rectangle (0.2cm,0.25cm); },
	axis on top,
	major grid style=white,
	ymajorgrids,
	extra x ticks={0.5,1.5,2.5,3.5,4.5,5.5,6.5},
    extra x tick labels={},
    extra x tick style={
        grid=major,
        major tick length=0pt,
		major grid style = {lightgray},
    },
        legend image post style={scale=0.5},
	legend style={draw=none,fill opacity=0.8,text opacity=1,/tikz/every even column/.append style={column sep=0.2cm}, legend columns=-1, font=\footnotesize}
]
\addplot table [x expr=\coordindex, y expr=\thisrow{realsize}/\thisrow{hadoop}, col sep=comma] {CSVs/wordcount-execution-time.csv};
\addplot table [x expr=\coordindex, y expr=\thisrow{realsize}/\thisrow{spark}, col sep=comma] {CSVs/wordcount-execution-time.csv};
\addplot table [x expr=\coordindex, y expr=\thisrow{realsize}/\thisrow{unicage}, col sep=comma] {CSVs/wordcount-execution-time.csv};
\legend{Hadoop MR,Spark,Unicage}
\end{axis}
\end{tikzpicture}
\centering
\caption{}
\label{fig:wordcount-dps}
\end{subfigure}
}
\caption{Data processing execution times (a) and rates (b) for \textit{wordcount}.}
\end{figure}
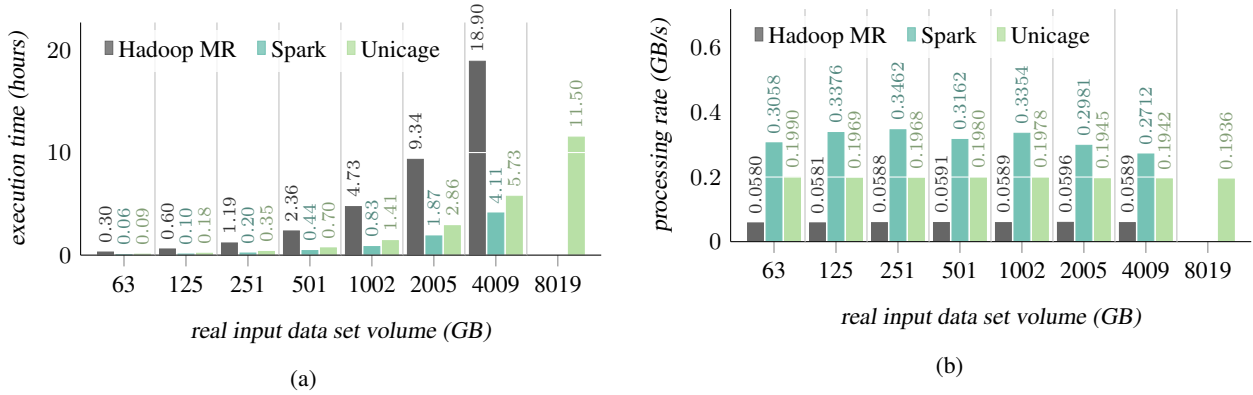

Figures~\ref{fig:wordcount-execution-time} and~\ref{fig:wordcount-dps} show the execution times and processing rates, respectively, of each system performing the \emph{wordcount} workload for the tested input data set volumes.

The performance of the \emph{wordcount} implementation with Unicage was superior to that of Hadoop and inferior to that of Spark for the data sets with volumes ranging from 63\,GB to 4009\,GB. The \emph{wordcount} implementation with Spark failed to produce correct outputs for the 8019\,GB data set, as the JVM ran out of space in the Java heap during the execution of the workload, as indicated by the throwing of a \texttt{java.lang.OutOfMemoryError} exception.

The output of the \emph{wordcount} implementation with Unicage for the 8019 GB input data set was verified through the comparison with the output of a single run of the \emph{wordcount} implementation with Hadoop, which took approximately 38 hours to complete, and thus was not validated statistically.

\subsection{Query Processing Benchmarks}

The structured e-commerce item/order tables data sets were used as input data sets for the query processing workloads in this benchmark set.

\subsubsection{Data Loading - Structured Data Set}

\begin{figure}[h]
\makebox[\textwidth][c]{
\begin{subfigure}{.515\textwidth}
\begin{tikzpicture}
\begin{axis}[
	ybar stacked,
    ybar = 1pt,
    bar width=6pt,
	xtick = {0,1,2,3,4,5,6,7},
	xticklabels = {68,139,281,565,1145,2320,4672,9417},
    x tick label style={
        /pgf/number format/1000 sep=, font=\small
    },
    y tick label style={
        font=\small
    },
    enlarge x limits = 0.1,
	width = \columnwidth,
	height = 0.63\columnwidth,
	xlabel = real input data set volume (GB),
	ylabel = execution time (hours),
	legend cell align = {left},
	legend pos = north west,
    xmin = 0,
    xmode = normal,
	ymajorgrids=true,
	major grid style = {lightgray},
	minor grid style = {lightgray!25},
	compat=newest,
	nodes near coords style={font=\scriptsize, rotate=90, anchor=west, /pgf/number format/.cd, fixed, fixed zerofill, precision=2, /tikz/.cd},
	enlarge y limits={upper, value=0.2},
	axis x line*=bottom,
	axis y line*=left,
	ymin=0,
	ytick style={draw=none},
	cycle list={
		{fill=black!80,draw=black!80,text=black!80},
		{fill=black!40,draw=black!40,text=black!60},
		{fill=black!40,draw=black!40,text=black!60},
		{fill=bargreen!100,draw=bargreen!100,text=bargreen!70!black},
		{fill=bargreen!100,draw=bargreen!100,text=bargreen!70!black}
	},
	legend image code/.code={
        \draw [#1] (0cm,-0.1cm) rectangle (0.2cm,0.25cm); },
	axis on top,
	major grid style=white,
	ymajorgrids,
	extra x ticks={0.5,1.5,2.5,3.5,4.5,5.5,6.5},
    extra x tick labels={},
    extra x tick style={
        grid=major,
        major tick length=0pt,
		major grid style = {lightgray},
    },
        legend image post style={scale=0.5},
	legend style={draw=none,fill opacity=0.8,text opacity=1,/tikz/every even column/.append style={column sep=0.2cm}, font=\footnotesize}
]
\addplot+[bar shift=-3pt] table [x expr=\coordindex, y expr=\thisrow{hdfs}/3600, col sep=comma] {CSVs/query-data-loading.csv};
\addplot+[hide axis, bar shift=-3pt] table [x expr=\coordindex, y expr=\thisrow{hive}/3600, col sep=comma] {CSVs/query-data-loading.csv};
\addplot+[hide axis, bar shift=-3pt, nodes near coords] coordinates {(0,0.000000001) (1,0.000000001) (2,0.000000001) (3,0.000000001) (4,0.000000001) (5,0.000000001) (6,0.000000001) (7,0.000000001)};
\resetstackedplots
\addplot+[hide axis, bar shift=+3pt] table [x expr=\coordindex, y expr=\thisrow{unicage}/3600, col sep=comma] {CSVs/query-data-loading.csv};
\addplot+[hide axis, bar shift=+3pt, nodes near coords] coordinates {(0,0.000000001) (1,0.000000001) (2,0.000000001) (3,0.000000001) (4,0.000000001) (5,0.000000001) (6,0.000000001) (7,0.000000001)};
\legend{Hadoop cluster - loading to HDFS,Hadoop cluster - creating Hive tables,,Unicage cluster,}
\end{axis}
\end{tikzpicture}
\centering
\caption{}
\label{fig:query-data-loading}
\end{subfigure}
\begin{subfigure}{.515\textwidth}
\begin{tikzpicture}
\begin{axis}[
    ybar = 1pt,
    bar width=6pt,
	xtick = {0,1,2,3,4,5,6,7},
	xticklabels = {68,139,281,565,1145,2320,4672,9417},
    x tick label style={
        /pgf/number format/1000 sep=, font=\small
    },
    y tick label style={
        font=\small
    },
    enlarge x limits = 0.1,
	width = \columnwidth,
	height = 0.63\columnwidth,
	xlabel = real input data set volume (GB),
	ylabel = loading rate (GB/s),
	legend cell align = {left},
	legend pos = north west,
    xmin = 0,
    xmode = normal,
	ymajorgrids=true,
	major grid style = {lightgray},
	minor grid style = {lightgray!25},
	compat=newest,
	nodes near coords,
	nodes near coords style={font=\scriptsize, rotate=90, anchor=west, /pgf/number format/.cd, fixed, fixed zerofill, precision=4, /tikz/.cd},
	enlarge y limits={upper, value=0.2},
	axis x line*=bottom,
	axis y line*=left,
	ymin=0,
	ymax=1.80,
	ytick style={draw=none},
	cycle list={
		{fill=black!80,draw=black!80,text=black!80},
		{fill=black!40,draw=black!40,text=black!60},
		{fill=bargreen!100,draw=bargreen!100,text=bargreen!70!black}
	},
	legend image code/.code={
        \draw [#1] (0cm,-0.1cm) rectangle (0.2cm,0.25cm); },
	axis on top,
	major grid style=white,
	ymajorgrids,
	extra x ticks={0.5,1.5,2.5,3.5,4.5,5.5,6.5},
    extra x tick labels={},
    extra x tick style={
        grid=major,
        major tick length=0pt,
		major grid style = {lightgray},
    },
        legend image post style={scale=0.5},
	legend style={draw=none,fill opacity=0.8,text opacity=1,/tikz/every even column/.append style={column sep=0.2cm}, legend columns=2, transpose legend, font=\scriptsize}
]
\addplot table [x expr=\coordindex, y expr=\thisrow{realsize}/\thisrow{hdfs}, col sep=comma] {CSVs/query-data-loading.csv};
\addplot table [x expr=\coordindex, y expr=\thisrow{realsize}/\thisrow{hive}, col sep=comma] {CSVs/query-data-loading.csv};
\addplot table [x expr=\coordindex, y expr=\thisrow{realsize}/\thisrow{unicage}, col sep=comma] {CSVs/query-data-loading.csv};
\legend{Hadoop cluster - loading to HDFS,Hadoop cluster - creating Hive tables,Unicage cluster}
\end{axis}
\end{tikzpicture}
\centering
\caption{}
\label{fig:query-data-loading-dps}
\end{subfigure}
}
\caption{Data loading execution times (a) and rates (b) for the e-commerce tables structured data set.}
\end{figure}
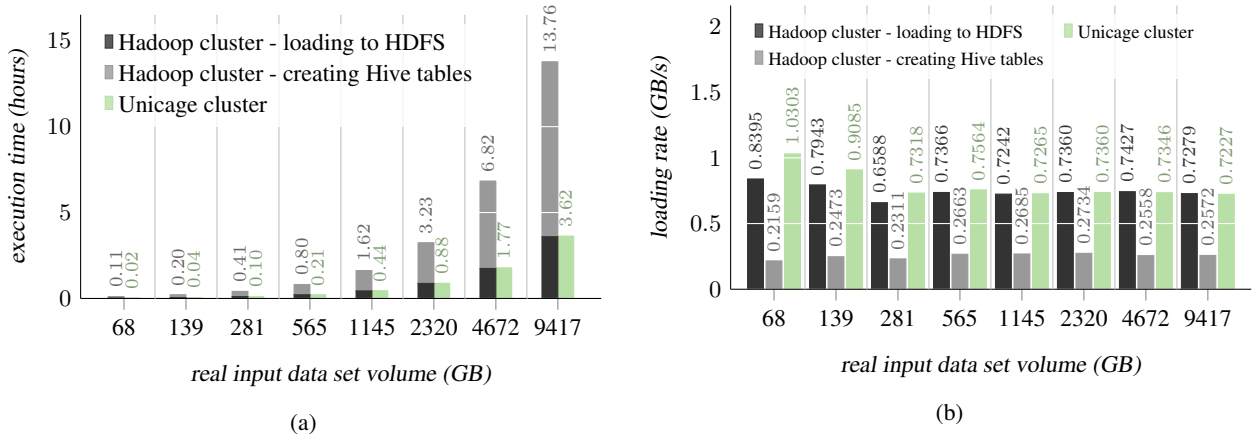

Figures~\ref{fig:query-data-loading} and~\ref{fig:query-data-loading-dps} show the execution times and loading rates, respectively, of loading the structured e-commerce item/order tables data sets into both the Hadoop and Unicage clusters, for the tested input data set volumes.

The stacked bar in Figure~\ref{fig:query-data-loading} indicates loading a structured data set into the Hadoop cluster required the step of loading it into HDFS, and the additional step of creating the Hive tables to support its schema. Loading the same data set into the Unicage cluster was significantly faster, as Unicage handles structured and unstructured data equally, as long as the data sets can be represented as text files in the file system. The trade-off is that in the absense of formal schema, it is of the responsibility of the developer to preserve the structure and veracity of the data set.

\subsubsection{Select}

\begin{figure}[h]
\makebox[\textwidth][c]{
\begin{subfigure}{.515\textwidth}
\begin{tikzpicture}
\begin{axis}[
    ybar = 1pt,
    bar width=6pt,
	xtick = {0,1,2,3,4,5,6,7},
	xticklabels = {43,86,173,347,698,1407,2823,5672},
    x tick label style={
        /pgf/number format/1000 sep=, font=\small
    },
    y tick label style={
        font=\small
    },
    enlarge x limits = 0.1,
	width = \columnwidth,
	height = 0.55\columnwidth,
	xlabel = real input data set volume (GB),
	ylabel = execution time (hours),
	legend cell align = {left},
	legend pos = north west,
    xmin = 0,
    xmode = normal,
	ymajorgrids=true,
	major grid style = {lightgray},
	minor grid style = {lightgray!25},
	compat=newest,
	nodes near coords,
	nodes near coords style={font=\scriptsize, rotate=90, anchor=west, /pgf/number format/.cd, fixed, fixed zerofill, precision=2, /tikz/.cd},
	enlarge y limits={upper, value=0.2},
	axis x line*=bottom,
	axis y line*=left,
	ymin=0,
	ytick style={draw=none},
	cycle list={
		{fill=black!60,draw=black!60,text=black!80},
		{fill=sparkblue,draw=sparkblue,text=sparkblue!70!black},
		{fill=bargreen!100,draw=bargreen!100,text=bargreen!70!black}
	},
	legend image code/.code={
        \draw [#1] (0cm,-0.1cm) rectangle (0.2cm,0.25cm); },
	axis on top,
	major grid style=white,
	ymajorgrids,
	extra x ticks={0.5,1.5,2.5,3.5,4.5,5.5,6.5},
    extra x tick labels={},
    extra x tick style={
        grid=major,
        major tick length=0pt,
		major grid style = {lightgray},
    },
        legend image post style={scale=0.5},
	legend style={draw=none,fill opacity=0.8,text opacity=1,/tikz/every even column/.append style={column sep=0.2cm}, legend columns=-1, font=\footnotesize}
]
\addplot table [x expr=\coordindex, y expr=\thisrow{hive}/3600, col sep=comma] {CSVs/select-execution-time.csv};
\addplot table [x expr=\coordindex, y expr=\thisrow{spark}/3600, col sep=comma] {CSVs/select-execution-time.csv};
\addplot table [x expr=\coordindex, y expr=\thisrow{unicage}/3600, col sep=comma] {CSVs/select-execution-time.csv};
\legend{Hive MR,Spark,Unicage}
\end{axis}
\end{tikzpicture}
\centering
\caption{}
\label{fig:select-execution-time}
\end{subfigure}
\begin{subfigure}{.515\textwidth}
\begin{tikzpicture}
\begin{axis}[
    ybar = 1pt,
    bar width=6pt,
	xtick = {0,1,2,3,4,5,6,7},
	xticklabels = {43,86,173,347,698,1407,2823,5672},
    x tick label style={
        /pgf/number format/1000 sep=, font=\small
    },
    y tick label style={
        font=\small
    },
    enlarge x limits = 0.1,
	width = \columnwidth,
	height = 0.55\columnwidth,
	xlabel = real input data set volume (GB),
	ylabel = processing rate (GB/s),
	legend cell align = {left},
	legend pos = north west,
    xmin = 0,
    xmode = normal,
	ymajorgrids=true,
	major grid style = {lightgray},
	minor grid style = {lightgray!25},
	compat=newest,
	nodes near coords,
	nodes near coords style={font=\scriptsize, rotate=90, anchor=west, /pgf/number format/.cd, fixed, fixed zerofill, precision=4, /tikz/.cd},
	enlarge y limits={upper, value=0.2},
	axis x line*=bottom,
	axis y line*=left,
	ymin=0,
	ymax=1.60,
	ytick style={draw=none},
	cycle list={
		{fill=black!60,draw=black!60,text=black!80},
		{fill=sparkblue,draw=sparkblue,text=sparkblue!70!black},
		{fill=bargreen!100,draw=bargreen!100,text=bargreen!70!black}
	},
	legend image code/.code={
        \draw [#1] (0cm,-0.1cm) rectangle (0.2cm,0.25cm); },
	axis on top,
	major grid style=white,
	ymajorgrids,
	extra x ticks={0.5,1.5,2.5,3.5,4.5,5.5,6.5},
    extra x tick labels={},
    extra x tick style={
        grid=major,
        major tick length=0pt,
		major grid style = {lightgray},
    },
        legend image post style={scale=0.5},
	legend style={draw=none,fill opacity=0.8,text opacity=1,/tikz/every even column/.append style={column sep=0.2cm}, legend columns=-1, font=\footnotesize}
]
\addplot table [x expr=\coordindex, y expr=\thisrow{realsize}/\thisrow{hive}, col sep=comma] {CSVs/select-execution-time.csv};
\addplot table [x expr=\coordindex, y expr=\thisrow{realsize}/\thisrow{spark}, col sep=comma] {CSVs/select-execution-time.csv};
\addplot table [x expr=\coordindex, y expr=\thisrow{realsize}/\thisrow{unicage}, col sep=comma] {CSVs/select-execution-time.csv};
\legend{Hive MR,Spark,Unicage}
\end{axis}
\end{tikzpicture}
\centering
\caption{}
\label{fig:select-dps}
\end{subfigure}
}
\caption{Data processing execution times (a) and rates (b) for \textit{select}.}
\end{figure}
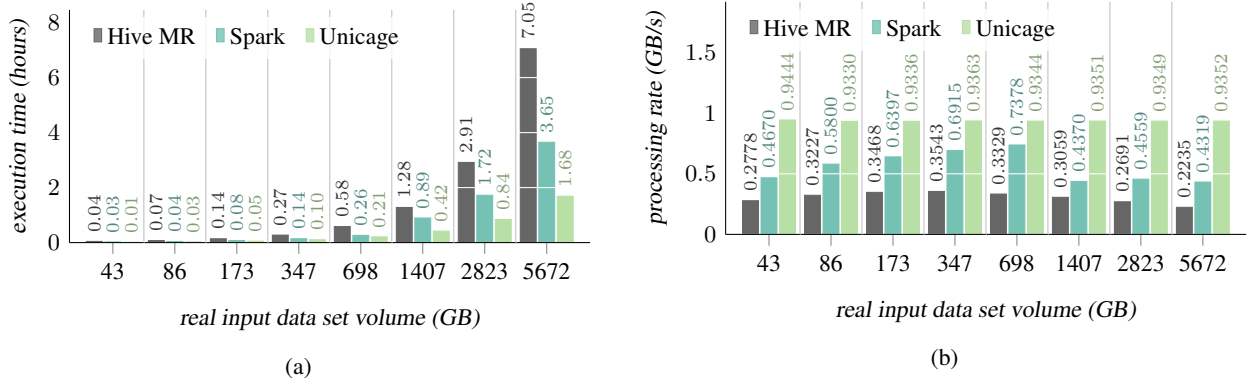

Figures~\ref{fig:select-execution-time} and~\ref{fig:select-dps} show the execution times and processing rates, respectively, of each big data system under test, when performing the \emph{select} workload for the tested input data set volumes.

The \emph{select} workload used the structured e-commerce tables, but only performed the search on the \emph{item} table. Therefore, the horizontal axis of the execution time and processing rate charts reflect the real volumes of the \emph{item} table only, as shown previously in Table~\ref{tab:query-gen-tables}.

The performance of the \emph{select} implementation with Unicage was superior to those of Hive and Spark, across all tested input data set volumes. 

\subsubsection{Join}

\begin{figure}[H]
\makebox[\textwidth][c]{
\begin{subfigure}{.515\textwidth}
\begin{tikzpicture}
\begin{axis}[
    ybar = 1pt,
    bar width=6pt,
	xtick = {0,1,2,3,4,5,6,7},
	xticklabels = {68,139,281,565,1145,2320,4672,9417},
    x tick label style={
        /pgf/number format/1000 sep=, font=\small
    },
    y tick label style={
        font=\small
    },
    enlarge x limits = 0.1,
	width = \columnwidth,
	height = 0.55\columnwidth,
	xlabel = real input data set volume (GB),
	ylabel = execution time (hours),
	legend cell align = {left},
	legend pos = north west,
    xmin = 0,
    xmode = normal,
	ymajorgrids=true,
	major grid style = {lightgray},
	minor grid style = {lightgray!25},
	compat=newest,
	nodes near coords,
	nodes near coords style={font=\scriptsize, rotate=90, anchor=west, /pgf/number format/.cd, fixed, fixed zerofill, precision=2, /tikz/.cd},
	enlarge y limits={upper, value=0.2},
	axis x line*=bottom,
	axis y line*=left,
	ymin=0,
	ytick style={draw=none},
	cycle list={
		{fill=black!60,draw=black!60,text=black!80},
		{fill=sparkblue,draw=sparkblue,text=sparkblue!70!black},
		{fill=bargreen!100,draw=bargreen!100,text=bargreen!70!black}
	},
	legend image code/.code={
        \draw [#1] (0cm,-0.1cm) rectangle (0.2cm,0.25cm); },
	axis on top,
	major grid style=white,
	ymajorgrids,
	extra x ticks={0.5,1.5,2.5,3.5,4.5,5.5,6.5},
    extra x tick labels={},
    extra x tick style={
        grid=major,
        major tick length=0pt,
		major grid style = {lightgray},
    },
        legend image post style={scale=0.5},
	legend style={draw=none,fill opacity=0.8,text opacity=1,/tikz/every even column/.append style={column sep=0.2cm}, legend columns=-1, font=\footnotesize}
]
\addplot table [x expr=\coordindex, y expr=\thisrow{hive}/3600, col sep=comma] {CSVs/join-execution-time.csv};
\addplot table [x expr=\coordindex, y expr=\thisrow{spark}/3600, col sep=comma] {CSVs/join-execution-time.csv};
\addplot table [x expr=\coordindex, y expr=\thisrow{unicage}/3600, col sep=comma] {CSVs/join-execution-time.csv};
\legend{Hive MR,Spark,Unicage}
\end{axis}
\end{tikzpicture}
\centering
\caption{}
\label{fig:join-execution-time}
\end{subfigure}
\begin{subfigure}{.515\textwidth}
\begin{tikzpicture}
\begin{axis}[
    ybar = 1pt,
    bar width=6pt,
	xtick = {0,1,2,3,4,5,6,7},
	xticklabels = {68,139,281,565,1145,2320,4672,9417},
    x tick label style={
        /pgf/number format/1000 sep=, font=\small
    },
    y tick label style={
        font=\small
    },
    enlarge x limits = 0.1,
	width = \columnwidth,
	height = 0.55\columnwidth,
	xlabel = real input data set volume (GB),
	ylabel = processing rate (GB/s),
	legend cell align = {left},
	legend pos = north west,
    xmin = 0,
    xmode = normal,
	ymajorgrids=true,
	major grid style = {lightgray},
	minor grid style = {lightgray!25},
	compat=newest,
	nodes near coords,
	nodes near coords style={font=\scriptsize, rotate=90, anchor=west, /pgf/number format/.cd, fixed, fixed zerofill, precision=4, /tikz/.cd},
	enlarge y limits={upper, value=0.2},
	axis x line*=bottom,
	axis y line*=left,
	ymin=0,
	ymax=0.40,
	ytick style={draw=none},
	cycle list={
		{fill=black!60,draw=black!60,text=black!80},
		{fill=sparkblue,draw=sparkblue,text=sparkblue!70!black},
		{fill=bargreen!100,draw=bargreen!100,text=bargreen!70!black}
	},
	legend image code/.code={
        \draw [#1] (0cm,-0.1cm) rectangle (0.2cm,0.25cm); },
	axis on top,
	major grid style=white,
	ymajorgrids,
	extra x ticks={0.5,1.5,2.5,3.5,4.5,5.5,6.5},
    extra x tick labels={},
    extra x tick style={
        grid=major,
        major tick length=0pt,
		major grid style = {lightgray},
    },
        legend image post style={scale=0.5},
	legend style={draw=none,fill opacity=0.8,text opacity=1,/tikz/every even column/.append style={column sep=0.2cm}, legend columns=-1, font=\footnotesize}
]
\addplot table [x expr=\coordindex, y expr=\thisrow{realsize}/\thisrow{hive}, col sep=comma] {CSVs/join-execution-time.csv};
\addplot table [x expr=\coordindex, y expr=\thisrow{realsize}/\thisrow{spark}, col sep=comma] {CSVs/join-execution-time.csv};
\addplot table [x expr=\coordindex, y expr=\thisrow{realsize}/\thisrow{unicage}, col sep=comma] {CSVs/join-execution-time.csv};
\legend{Hive MR,Spark,Unicage}
\end{axis}
\end{tikzpicture}
\centering
\caption{}
\label{fig:join-dps}
\end{subfigure}
}
\caption{Data processing execution times (a) and rates (b) for \textit{join}.}
\end{figure}
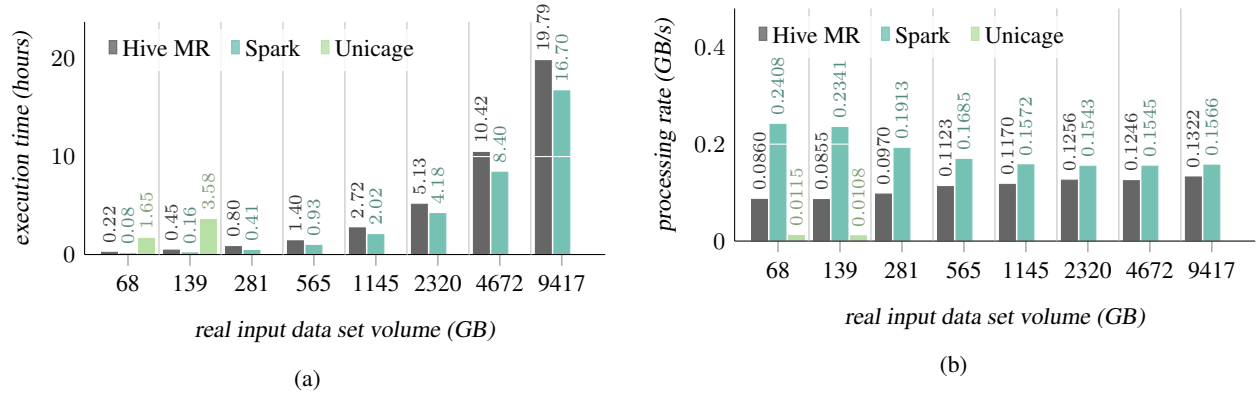

Figures~\ref{fig:join-execution-time} and~\ref{fig:join-dps} show the execution times and processing rates, respectively, of each system under test, when performing the \emph{join} workload for the input data set volumes.

The performance of the \emph{join} implementation with Unicage was vastly inferior to those of Hive and Spark for the 68\,GB and 139\,GB input data sets, and produced arbitrarily incorrect outputs for the 281\,GB and larger data sets.

\subsubsection{Aggregation}

\begin{figure}[h]
\makebox[\textwidth][c]{
\begin{subfigure}{.515\textwidth}
\begin{tikzpicture}
\begin{axis}[
    ybar = 1pt,
    bar width=6pt,
	xtick = {0,1,2,3,4,5,6,7},
	xticklabels = {43,86,173,347,698,1407,2823,5672},
    x tick label style={
        /pgf/number format/1000 sep=, font=\small
    },
    y tick label style={
        font=\small
    },
    enlarge x limits = 0.1,
	width = \columnwidth,
	height = 0.55\columnwidth,
	xlabel = real input data set volume (GB),
	ylabel = execution time (hours),
	legend cell align = {left},
	legend pos = north west,
    xmin = 0,
    xmode = normal,
	ymajorgrids=true,
	major grid style = {lightgray},
	minor grid style = {lightgray!25},
	compat=newest,
	nodes near coords,
	nodes near coords style={font=\scriptsize, rotate=90, anchor=west, /pgf/number format/.cd, fixed, fixed zerofill, precision=2, /tikz/.cd},
	enlarge y limits={upper, value=0.2},
	axis x line*=bottom,
	axis y line*=left,
	ymin=0,
	ytick style={draw=none},
	cycle list={
		{fill=black!60,draw=black!60,text=black!80},
		{fill=sparkblue,draw=sparkblue,text=sparkblue!70!black},
		{fill=bargreen!100,draw=bargreen!100,text=bargreen!70!black}
	},
	legend image code/.code={
        \draw [#1] (0cm,-0.1cm) rectangle (0.2cm,0.25cm); },
	axis on top,
	major grid style=white,
	ymajorgrids,
	extra x ticks={0.5,1.5,2.5,3.5,4.5,5.5,6.5},
    extra x tick labels={},
    extra x tick style={
        grid=major,
        major tick length=0pt,
		major grid style = {lightgray},
    },
        legend image post style={scale=0.5},
	legend style={draw=none,fill opacity=0.8,text opacity=1,/tikz/every even column/.append style={column sep=0.2cm}, legend columns=-1, font=\footnotesize}
]
\addplot table [x expr=\coordindex, y expr=\thisrow{hive}/3600, col sep=comma] {CSVs/aggregation-execution-time.csv};
\addplot table [x expr=\coordindex, y expr=\thisrow{spark}/3600, col sep=comma] {CSVs/aggregation-execution-time.csv};
\addplot table [x expr=\coordindex, y expr=\thisrow{unicage}/3600, col sep=comma] {CSVs/aggregation-execution-time.csv};
\legend{Hive MR,Spark,Unicage}
\end{axis}
\end{tikzpicture}
\centering
\caption{}
\label{fig:aggregation-execution-time}
\end{subfigure}
\begin{subfigure}{.515\textwidth}
\begin{tikzpicture}
\begin{axis}[
    ybar = 1pt,
    bar width=6pt,
	xtick = {0,1,2,3,4,5,6,7},
	xticklabels = {43,86,173,347,698,1407,2823,5672},
    x tick label style={
        /pgf/number format/1000 sep=, font=\small
    },
    y tick label style={
        font=\small
    },
    enlarge x limits = 0.1,
	width = \columnwidth,
	height = 0.55\columnwidth,
	xlabel = real input data set volume (GB),
	ylabel = processing rate (GB/s),
	legend cell align = {left},
	legend pos = north west,
    xmin = 0,
    xmode = normal,
	ymajorgrids=true,
	major grid style = {lightgray},
	minor grid style = {lightgray!25},
	compat=newest,
	nodes near coords,
	nodes near coords style={font=\scriptsize, rotate=90, anchor=west, /pgf/number format/.cd, fixed, fixed zerofill, precision=4, /tikz/.cd},
	enlarge y limits={upper, value=0.2},
	axis x line*=bottom,
	axis y line*=left,
	ymin=0,
	ymax=0.90,
	ytick style={draw=none},
	cycle list={
		{fill=black!60,draw=black!60,text=black!80},
		{fill=sparkblue,draw=sparkblue,text=sparkblue!70!black},
		{fill=bargreen!100,draw=bargreen!100,text=bargreen!70!black}
	},
	legend image code/.code={
        \draw [#1] (0cm,-0.1cm) rectangle (0.2cm,0.25cm); },
	axis on top,
	major grid style=white,
	ymajorgrids,
	extra x ticks={0.5,1.5,2.5,3.5,4.5,5.5,6.5},
    extra x tick labels={},
    extra x tick style={
        grid=major,
        major tick length=0pt,
		major grid style = {lightgray},
    },
        legend image post style={scale=0.5},
	legend style={draw=none,fill opacity=0.8,text opacity=1,/tikz/every even column/.append style={column sep=0.2cm}, legend columns=-1, font=\footnotesize}
]
\addplot table [x expr=\coordindex, y expr=\thisrow{realsize}/\thisrow{hive}, col sep=comma] {CSVs/aggregation-execution-time.csv};
\addplot table [x expr=\coordindex, y expr=\thisrow{realsize}/\thisrow{spark}, col sep=comma] {CSVs/aggregation-execution-time.csv};
\addplot table [x expr=\coordindex, y expr=\thisrow{realsize}/\thisrow{unicage}, col sep=comma] {CSVs/aggregation-execution-time.csv};
\legend{Hive MR,Spark,Unicage}
\end{axis}
\end{tikzpicture}
\centering
\caption{}
\label{fig:aggregation-dps}
\end{subfigure}
}
\caption{Data processing execution times (a) and rates (b) for \textit{aggregation}.}
\end{figure}
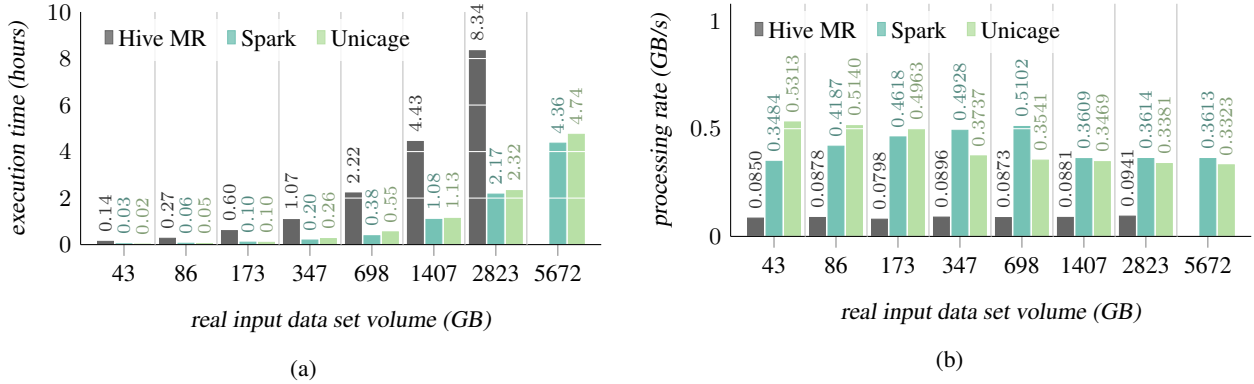

Figures~\ref{fig:aggregation-execution-time} and~\ref{fig:aggregation-dps} show the execution times and processing rates for the systems performing the \emph{aggregation} workload.

Similarly to the \emph{select} workload, the \emph{aggregation} workload only performed the grouping on the \emph{item} table. The horizontal axis of the execution time and processing rate charts reflect the real volumes of this table.
The performance of the \emph{aggregation} implementation with Unica\-ge was superior to those of Hive and Spark for the 63\,GB, 86\,GB and 173\,GB data sets, but inferior to that of Spark for the 347\,GB and larger data sets.

Many Unicage commands, including the \texttt{sm2} grouping operation used in the \emph{aggregation} implementation, require the input files to be sorted. Sorting with Unicage is most efficient with the \texttt{msort} command, which performs in-memory sorting. Each file of the structured e-commerce item/order tables data set can grow quite large, as shown previously in Table~\ref{tab:number_and_volume_files}. This caused the premature failure of the \emph{aggregation} implementation with Unicage, as \texttt{msort} ran out of memory for the 347\,GB data set. To overcome this, the implementation was refactored to split each input file in chunks small enough to be sorted by \texttt{msort}. This allowed the benchmarks to proceed, but caused the performance degradation of the \emph{aggregation} implementation with Unicage, observed for the 347\,GB and larger data sets.

The \emph{aggregation} implementation with Hive failed to produce correct outputs for the 5672\,GB data set, with a failing MapReduce stage. 

\subsection{Discussion}

Complex big data systems offer a plethora of abstractions to ease the implementation and execution of big data processing workloads. In our experimental setup, HDFS provided distributed storage and YARN provided distributed resource management to the Hadoop cluster. Additionally, in the Hadoop cluster, the Metastore of Hive formalized the structure of structured data sets.

\subsubsection{Distributed File System}

A sorted data set has inter-record dependencies, meaning each record has a placement dependency on other records. Unicage requires output data sets with this property to be stored in a single machine. For this reason, even though it failed prematurely, the \emph{sort} implementation with Unicage was predicted to fail for data sets larger than 1024 GB, since that was the maximum storage capacity of the \emph{unicageleader} node (see Table~\ref{tab:cluster}), where the outputs of the workloads were consolidated. 
In this aspect, Unicage can scale horizontally to improve the performance of the workloads, but is hindered by the capacity of singular machines, being highly dependent on vertical scaling to address larger volumes. 
HDFS offers the abstraction of a global namespace not only to the user but also to the applications, allowing large inter-record dependent outputs to be stored across multiple nodes. This is not immediately possible in lean big data systems without the integration of additional systems to support a similar abstraction to that of a distributed file system.

A similar comment can be made regarding the \emph{join} workload, which explores the inter-record dependencies of the input data set. To join two records from different tables, Unicage required these to be stored in the same node. As such, the \emph{join} implementation with Unicage required intermediate files to be moved around the cluster, severely crippling the performance. Since HDFS allows applications to access the distributed file system as a single global namespace, having joinable records in different machines was not a challenge for the tested complex big data systems.

\subsubsection{Distributed Resource Manager}

Another challenge of lean big data systems was the optimization of the resource usage, which was a manual endeavor that was only possible with the assistance of observability tools. This process was also error prone, as the resource usage patterns of shell script implementations often became difficult to predict with the increase in volume of the input data sets. Optimizing resource usage was not a challenge with the tested complex big data systems, as that is the main function of YARN.

\subsubsection{Structured Metadata Store}

Unicage requires data sets to be represented as text files to be processed with shell scripting.
Loading a structured data set with Unicage was faster because Unicage does not require data to be normalized in a database, as it delegates the responsibility of maintaining the structure of the data to the developer. 
This is not ideal when the structural schema is too elaborate, as the developer might compromise the veracity of the data set.  
In this context, complex systems with abstractions to formalize the structure of the data, like the Metastore of Hive, should be used.

\begin{figure}[ht]
\includegraphics[width=0.8\columnwidth]{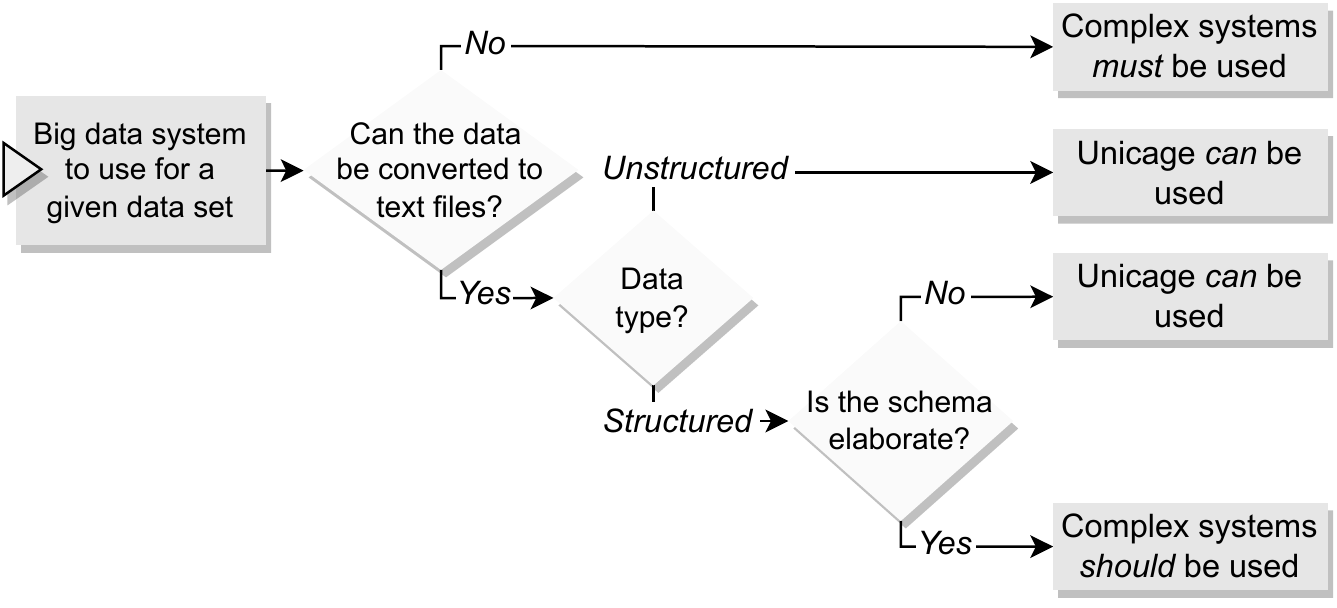}
\centering
\caption[Summary of the varieties of data Unicage can be used with.]{Summary of the varieties of data Unicage can be used with.}
\label{fig:varieties-breakdown}
\end{figure}

The previous statements are summarized in Figure~\ref{fig:varieties-breakdown}, which presents a flowchart highlighting the varieties of data for which Unicage \emph{can} be used; those for which Unicage is not ideal, and thus a complex system \emph{should} be used; and those for which Unicage cannot be used, and thus a complex system \emph{must} be used.

\subsubsection{Data Loading Performance - the Impact of Data Sets with Different File System Layouts}

The Wikipedia text entries data set was comprised of a growing number of 500\,MB files, while the e-commerce item/order tables data set was comprised of a fixed number of 8 files with growing volumes. This was previously highlighted in Table~\ref{tab:number_and_volume_files}. This dichotomy translated into a small performance impact for both data loading into the Hadoop cluster with HDFS and into the Unicage cluster. 

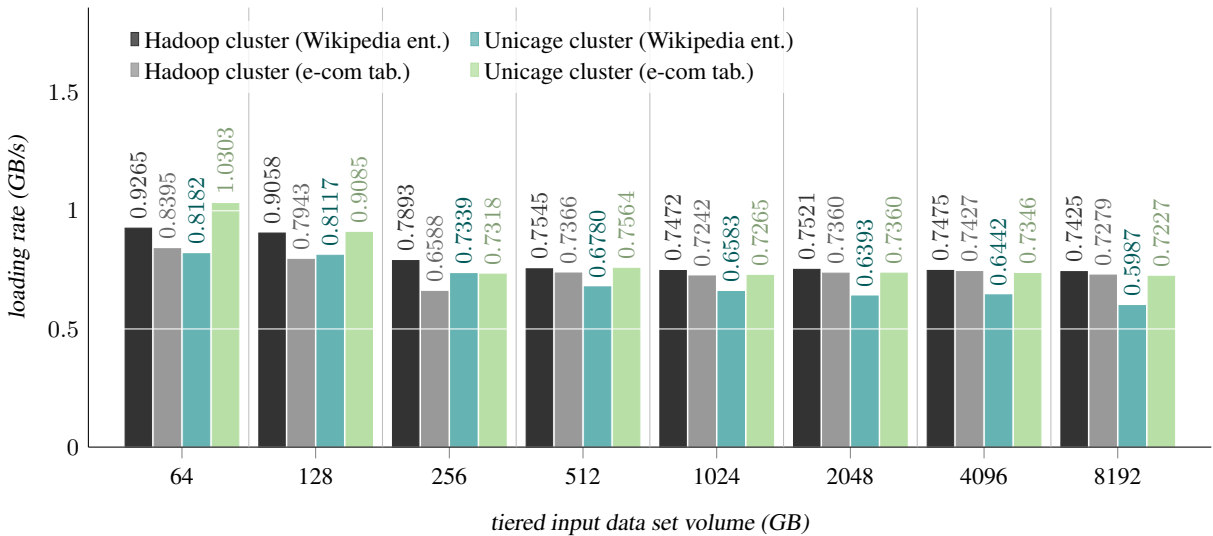
\begin{figure}[h]
\begin{tikzpicture}
\begin{axis}[
    ybar = 1pt,
    bar width=10pt,
	xtick = {0,1,2,3,4,5,6,7},
	xticklabels = {64,128,256,512,1024,2048,4096,8192},
    x tick label style={
        /pgf/number format/1000 sep=, font=\small
    },
    y tick label style={
        font=\small
    },
    enlarge x limits = 0.1,
	width = \columnwidth,
	height = 0.45\columnwidth,
	xlabel = tiered input data set volume (GB),
	ylabel = loading rate (GB/s),
	legend cell align = {left},
	legend pos = north west,
    xmin = 0,
    xmode = normal,
	ymajorgrids=true,
	major grid style = {lightgray},
	minor grid style = {lightgray!25},
	compat=newest,
	nodes near coords,
	nodes near coords style={font=\footnotesize, rotate=90, anchor=west, /pgf/number format/.cd, fixed, fixed zerofill, precision=4, /tikz/.cd},
	enlarge y limits={upper, value=0.2},
	axis x line*=bottom,
	axis y line*=left,
	ymin=0,
	ymax=1.55,
	ytick style={draw=none},
	cycle list={
		{fill=black!80,draw=black!80,text=black!80},
		{fill=black!40,draw=black!40,text=black!60},	{fill=teal!60,draw=teal!60,text=teal!70!black},
        {fill=bargreen!100,draw=bargreen!100,text=bargreen!70!black}
	},
	legend image code/.code={
        \draw [#1] (0cm,-0.1cm) rectangle (0.2cm,0.25cm); },
	axis on top,
	major grid style=white,
	ymajorgrids,
	extra x ticks={0.5,1.5,2.5,3.5,4.5,5.5,6.5},
    extra x tick labels={},
    extra x tick style={
        grid=major,
        major tick length=0pt,
		major grid style = {lightgray},
    },
        legend image post style={scale=0.5},
	legend style={draw=none,fill opacity=0.8,text opacity=1,/tikz/every even column/.append style={column sep=0.2cm}, legend columns=2, transpose legend, font=\footnotesize}
]
\addplot table [x expr=\coordindex, y=HDP-Wiki, col sep=comma] {CSVs/all-loading-rates.csv};
\addplot table [x expr=\coordindex, y=HDP-Ecom, col sep=comma] {CSVs/all-loading-rates.csv};
\addplot table [x expr=\coordindex, y=Unic-Wiki, col sep=comma] {CSVs/all-loading-rates.csv};
\addplot table [x expr=\coordindex, y=Unic-Ecom, col sep=comma] {CSVs/all-loading-rates.csv};
\legend{Hadoop cluster (Wikipedia ent.),Hadoop cluster (e-com tab.),Unicage cluster (Wikipedia ent.),Unicage cluster (e-com tab.)}
\end{axis}
\end{tikzpicture}
\centering
\caption[Composite chart of data loading rates.]{Composite chart of data loading rates.}
\label{fig:data-loading-dps}
\end{figure}

Figure~\ref{fig:data-loading-dps} shows a composite chart of the loading rates of all benchmarked data loading workloads. The vertical axis shows the loading rate, in gigabytes per second, and the horizontal axis shows the tiered input data set volumes, in gigabytes. For the structured e-commerce item/order tables data set, the process of creating the Hive tables in the Hadoop cluster was excluded from the chart because that is a consequence of the structural properties of the data set, and not of its layout in the file system, and as far as HDFS is aware, both data sets are just files. 

The performance of loading data into the Hadoop cluster with the \texttt{put} operation of HDFS is slightly increased for data sets with a large number of small files, such as the Wikipedia text entries data sets, providing each file is large enough to fill file blocks in the file system (128\,MB each, by default). The performance of loading data into the Unicage cluster with the \texttt{distr-distr} Unicage command is slightly increased for data sets with a small number of large files, such as the e-commerce item/order tables data sets.

\section{Conclusion}
\label{sec:conclusion}

\begin{figure}[ht]
\includegraphics[width=0.8\columnwidth]{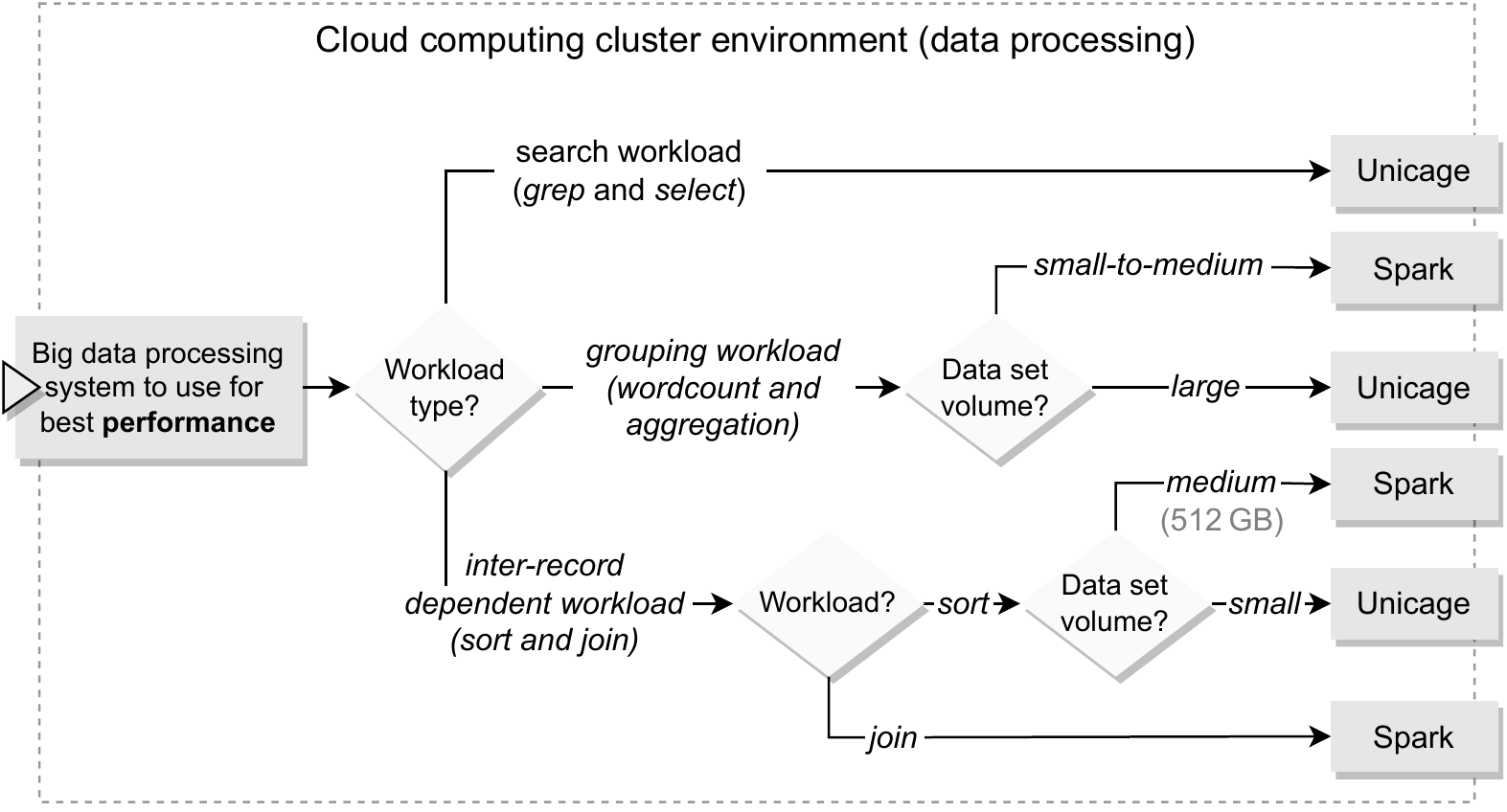}
\centering
\caption[Summary of conclusions from the data processing benchmarks.]{Summary of conclusions from the data processing benchmarks.}
\label{fig:benchmarks-breakdown}
\end{figure}

This paper presented a performance benchmark of two systems with two distinct approaches to big data processing in cluster environments: Spark using a complex software stack, and Unicage shell scripts using a lean software stack. 
We benchmarked the performance of batch processing workloads over unstructured data sets, and query processing workloads over structured data sets. The tiered volumes of the data sets ranged from 64\,GB to 8192\,GB. 
Figure~\ref{fig:benchmarks-breakdown} summarizes the results of these benchmarks.
The performance of data loading of structured and unstructured data was also benchmarked for the aforementioned data sets.

The decision between taking the complex or lean approaches will always depend on
the use-case, with multiple weighting factors, including the types of workloads, the volumes and
varieties of the data sets, and the available resources in the computing clusters. However, this study shows that, in terms of performance, there are big data workloads that do not require the use of complex big data systems:
for search workloads (\textit{grep} and \textit{select}), Unicage shell scripts were preferred to Spark, across all tested input data set volumes;
for grouping workloads (\textit{wordcount} and \textit{aggregation}), Unicage was only preferred to Spark for large data sets (8192\,GB);
for the \textit{sort} workload, Unicage was only preferred to Spark for smaller data sets (from 64\,GB to 256\,GB); 
for the \textit{join} workload, Spark was preferred to Unicage, across all tested input data set volumes.
The results of the \emph{sort} and \emph{join} workloads indicate that the used version of Unicage has a software defect that causes arbitrary faults in the output, with no crash report.
Even though Unicage produces logs of activity, the lack of an explicit fault hinting at incorrect outputs might be problematic in use-cases where the correctness of the outputs of the big data workloads is heavily relied upon.

Regarding the data loading, we have shown that the performance of loading unstructured data into a Unicage cluster was slightly inferior to that of loading the same data into a Hadoop cluster with HDFS, and the performance of loading structured data into a Unicage cluster was superior to that of loading the same data into a Hadoop cluster with HDFS and Hive, across all tested input data set volumes.
The trade-off is that Unicage does not normalize structured data in a database format, so the developer is charged with the responsibility of maintaining that structure. 

Our findings favor further work on complex and lean big data systems.
Lean systems can be used to do simple pre-processing (e.g., search workloads), reducing the size of the data sets before sending them over to complex systems like Spark, where more complex operations are performed, thus reducing overall costs. 
This can be especially relevant if combined with edge computing, where the edge nodes with limited computational and power resources are able to run lean software stacks but not complex ones. 
This possibility should be validated by future studies addressing end-to-end benchmarks with more complex sets of workloads in 
application-specific scenarios. 
Other types of data processing can also be benchmarked, including stream processing and machine learning. 
The obstacles found in the \textit{sort} and \textit{join} workloads motivate the future integration of 
distributed storage and 
resource usage abstractions with lean big data systems. The integration of recent technologies such as \textsc{PaSh}~\cite{vasilakis2021pash} and \textsc{Posh}~\cite{raghavan2020posh} with lean implementations could also provide further 
performance optimizations to the shell script-based approach.

\section*{Acknowledgements}
\label{sec:ack}

This research was possible thanks to Unicage, for providing complimentary access to their proprietary Unicage Tukubai and BOA libraries; and to IBM, for providing complimentary access to the IBM Cloud VPC infrastructure.


\bibliographystyle{unsrt}  
\bibliography{paper}

\end{document}